\documentclass[iop]{emulateapj}
\usepackage{amsmath}
\usepackage{hyperref}
\shorttitle{Photometric Spin-Orbit Misalignment}
\shortauthors{G. Li and J. Winn }

\begin{document}
\newcommand{\icarus}{Icarus}

\title{Are Tidal Effects Responsible for Exoplanetary Spin-Orbit Alignment?}
\author{Gongjie Li \altaffilmark{1} and Joshua N.\ Winn \altaffilmark{2}}
\affil{1. Harvard-Smithsonian Center for Astrophysics}
\affil{2. Department of Physics, Massachusetts Institute of Technology, 77 Massachusetts Avenue,
Cambridge, MA 02139}
\email{gli@cfa.harvard.edu}


\altaffiltext{1}{1Harvard-Smithsonian Center for Astrophysics, The Institute for Theory and
Computation, 60 Garden Street, Cambridge, MA 02138, USA}
\altaffiltext{2}{2Department of Physics, Massachusetts Institute of Technology,
Cambridge, Massachusetts, 02139, USA}


\begin{abstract}

  The obliquities of planet-hosting stars are clues about the formation of planetary systems. Previous observations led to the hypothesis that for close-in giant planets, spin-orbit alignment is enforced by tidal interactions. Here, we examine two problems with this hypothesis. First, Mazeh and coworkers recently used a new technique -- based on the amplitude of starspot-induced photometric variability -- to conclude that spin-orbit alignment is common even for relatively long-period planets, which would not be expected if tides were responsible. We re-examine the data and find a statistically significant correlation between photometric variability and planetary orbital period that is qualitatively consistent with tidal interactions. However it is still difficult to explain quantitatively, as it would require tides to be effective for periods as long as tens of days. Second, Rogers and Lin argued against a particular theory for tidal re-alignment by showing that initially retrograde systems would fail to be re-aligned, in contradiction with the observed prevalence of prograde systems. We investigate a simple model that overcomes this problem by taking into account the dissipation of inertial waves and the equilibrium tide, as well as magnetic braking. We identify a region of parameter space where re-alignment can be achieved, but it only works for close-in giant planets, and requires some fine tuning. Thus, while we find both problems to be more nuanced than they first appeared, the tidal model still has serious shortcomings.

\end{abstract}


\section{Introduction}

One surprising aspect of the geometry of exoplanetary systems is that the rotation of the host star can be drastically misaligned with the planetary orbits [see, e.g., \citet{Hebrard08, Winn09, Narita09, Triaud10, Hirano11, Albrecht12, Huber13, Li14_56}, or the review by \citet{Winn14}]. Many theories have been advanced to explain the misalignments. In some theories, the planet's orbit is tilted through few-body gravitational dynamics. This includes planet-planet scattering, as well as long-term secular dynamical effects between planets or involving a stellar companion \citep[see, e.g.,][]{Fabrycky07, Chatterjee08, Nagasawa08, Naoz11, Naoz12, Wu11, Li14, Valsecchi14}. In other theories the misalignment arises not between planets, but rather between the host star and the protoplanetary disk, or a tightly coupled system of coplanar planetary orbits. For example the star can be tilted relative to the protoplanetary disk through magnetic \citep{Lai11} or fluid-dynamical effects \citep{Rogers12,Rogers13}. Alternatively, the disk can be tilted because of the inhomogeneity of the collapse of the original molecular cloud \citep{Bate10, Fielding15}, or the gravitational torque from a passing star \citep{Tremaine89, Thies11, Batygin12}.

One seemingly important clue is that for host stars of hot Jupiters -- the most thoroughly investigated type of system -- misalignments are seen more frequently among relatively hot stars ($T_{\rm eff}\gtrsim 6100$~K) than cooler stars \citep{Schlaufman10, Winn10, Albrecht12}.  Another possible clue, though with less secure support, is that the host stars of the most massive hot Jupiters ($\gtrsim$3~$M_{\rm Jup}$) tend to have lower obliquities \citep{Hebrard11, Dawson14}. One possible explanation for these trends invokes the coplanarizing action of star-planet tides \citep{Winn10}, which is thought to be more rapid for cooler stars than for hotter stars, owing to differences in their internal structures. The stronger magnetic braking of the cool stars may also play an important role \citep{Dawson14}. Thus, one hypothesis for the spin-orbit misalignments that is consistent with all the preceding results is that they are a frequent by-product of dynamical effects, but the misalignments are eventually erased if tidal dissipation and magnetic braking are sufficiently strong. Recently this hypothesis has been criticized, on both observational and theoretical grounds.

On the observational side, \citet{Mazeh15} found that the differing obliquity distributions of hot and cool stars seems to exist even for relatively long-period planets, whose orbits are too distant for tidal interactions to be relevant.  To arrive at this result, they applied a novel technique to the {\it Kepler} sample of transiting planets.  Their technique was based on the observed level of photometric variability of the host star due to starspots carried around by stellar rotation. All other things being equal, when the star's rotation axis is viewed at high inclination, the photometric variability should be larger than when it is viewed at low inclination.  Since the orbit of a transiting planet is always viewed at high inclination, a population of transit-hosting stars with low obliquities should show systematically greater photometric variability than a population of randomly oriented stars. Indeed, \citet{Mazeh15} found this effect for relatively cool stars ($T_{\rm eff} \lesssim 6000$~K), whereas the hotter stars were more consistent with random orientations. And, most pertinent to this paper, \citet{Mazeh15} found that the enhanced variability of cool stars did not seem to depend on the orbital period of the planet, all the way out to the $\approx$50-day limit of the sample.  In contrast, under the tidal re-alignment hypothesis, one would expect the obliquity distribution to depend on period, with lower obliquities for the closest-in planets.  For a Sun-like star in particular, a period of 50~days corresponds to an orbital distance of about 0.25~AU, which is thought to be too large for tidal interactions to be relevant.

On the theoretical side, from the moment the tidal hypothesis was made, it was recognized that it may be difficult to realign the system without also destroying the planet through orbital decay.  The very simplest tidal theories (involving only the dissipation of the equilibrium tide) predict similar timescales for the spin-up of the star, the alignment of the star and orbit, and the shrinkage of the orbital distance.  \citet{Winn10} suggested that the problem could be overcome if only the star's outer convective layer participates in the re-alignment, but it seems unlikely that the interior could remain uncoupled for a sufficiently long time. \citet{Lai12} rescued the hypothesis, by identifying a component in the tidal torque of a misaligned system which can reduce the obliquity without causing orbital decay.  However, \citet{Rogers13b} have criticized this theory, by showing that this same component of the tidal potential should lead with substantial probability to final states in which the spin and orbit are retrograde or perpendicular. This does not agree with the observed preponderance of prograde configurations. \citet{Xue14} pointed out that when the effects of both the equilibrium tide and inertial waves are taken into account,
re-alignment can be achieved before orbital decay. Here we study the parameter space of this type of scenario, including the equilibrium tide, inertial waves and magnetic braking.

In this paper we examine these two problems in more detail.  In~\textsection\ref{s:var}, we re-examine the technique and the dataset presented by \citet{Mazeh15} and search for any statistical evidence for a period-dependence of the obliquity distribution.  In~\textsection\ref{s:tide}, we investigate simple tidal models including the effects of the equilibrium tide, magnetic braking, and the tidal torque component identified by \citet{Lai12}, to see if it is possible for the final states to be predominantly prograde systems.  In~\textsection\ref{s:diss} we summarize our conclusions and their implications for theories of planetary formation and evolution.

\section{Tests for a Period-Dependence of Photometric Variability}
\label{s:var}

The basis for the pioneering work of \citet{Mazeh15} was the collection and update of rotation periods and associated amplitudes of photometric variability that was presented by \citet{McQuillan13} and \citet{McQuillan14}, based on {\it Kepler} data. The sample includes 34030 main-sequence stars, and 993 stars that are identified as likely hosts of transiting planets; the latter are referred to as {\it Kepler} Objects of Interest or KOIs.  The photometric variability amplitude was quantified by the statistic $R_{\rm var}$, defined as the amplitude of the photometric modulation in parts per million. They compared the distributions of $R_{\rm var}$ between different samples of stars. For the relatively cool stars ($T_{\rm eff} < 6000$~K), they found that $R_{\rm var}$ tends to be higher for the KOIs than the non-KOIs.  In contrast, for the hotter stars, they found that $R_{\rm var}$ tends to be lower for the KOIs than the non-KOIs. They also divided the KOIs into two groups, short-period (1-5~days) and long-period (5-50~days), and did not identify any systematic difference in $R_{\rm var}$ between these groups. In this section we re-examine the same dataset, looking specifically for any period dependence of $R_{\rm var}$ among the sample of cool KOIs.

We obtain the data for $R_{\rm var}$, $P_{\rm rot}$, and $T_{\rm eff}$ from \citet{Mazeh15}. For the KOIs, we also obtain the corresponding transit-related data (stellar radius, orbital periods, transit signal-to-noise ratio, and root-mean-squared average of the combined differential photometric precision) from the NASA Exoplanet Archive\footnote{\url{http://exoplanetarchive.ipac.caltech.edu/}, queried by 2015~May~21} (\citealt{Akeson13}, NEA). In particular, the stellar radii in the NEA were taken from the compilation of \citet{Huber14}.  We exclude targets which are identified as false positives in the NEA.  We also exclude six KOIs for which no stellar radius is given in the NEA, and a single KOI (1174) for which the NEA lists a mysteriously long orbital period of $1.3\times10^5$ days. We are left with 909 KOIs. Of these, 814 have $T_{\rm eff} < 6000$~K and are the focus of this paper; henceforth, for brevity, we refer to this sample as the KOIs.  (The sample of hotter KOIs is too small for a meaningful search for period-dependent effects.)  As a control sample, we use the $R_{\rm var}$ data for 34030 main-sequence stars with $T_{\rm eff} < 6000$~K given by \citet{McQuillan14}; we will refer to this sample as the ``non-KOIs''.

\subsection{Statistical tests and results}

\subsubsection{Correlation Coefficients}

\begin{figure}
\includegraphics[width=3.2in]{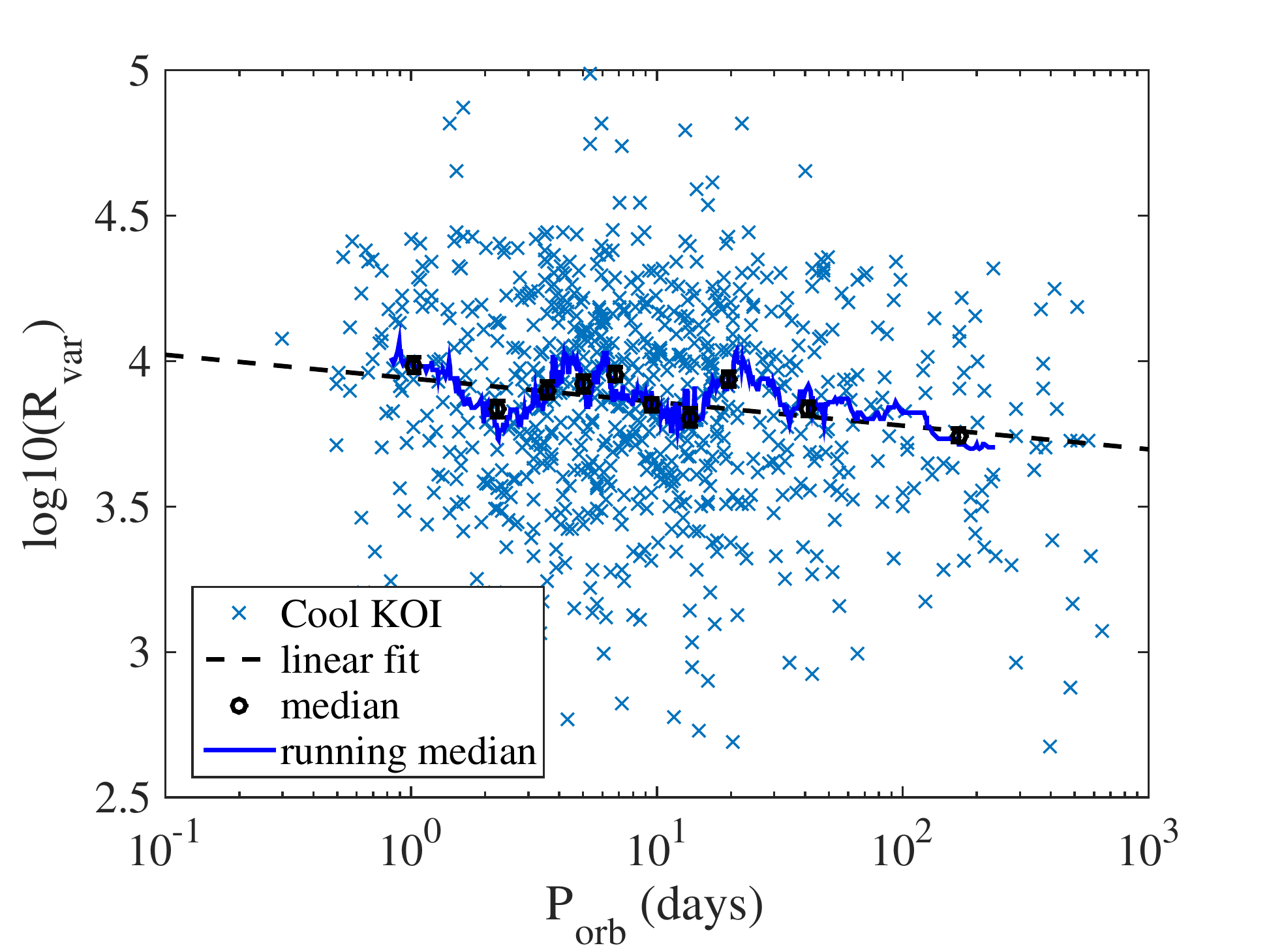}
\caption{Observed decrease in photometric variability amplitude as a function of planetary orbital period, for the 814 cool KOIs ($T_{\rm eff}< 6000K$). The black dashed line represents the linear fit. The black circles show the median $R_{\rm var}$ within ten period bins. The blue line is the running median, using a smoothing width of 50 data points. The uncertainties in the median of the ten equally sized bins are around $0.026$ (as shown in this figure) and those of the running median are around $0.033$. }
\label{f:lmfit}
\end{figure}

To study the overall dependence of the photometric variability on the planet's orbital period, we perform a linear regression of the variability amplitude of the KOIs and orbital period:
\begin{equation}
\label{e:linear}
\log{R_{\rm var}} = a_0 + a_1 \log{P_{\rm orb}} + \epsilon \,
\end{equation}
where $R_{\rm var}$ is the variability amplitude in parts per million, and $P_{\rm orb}$ is the orbital period of the planet in days. For multi-transiting systems, we use the orbital period of the innermost planet.
(Comparison of the multi-transiting systems and the single-transiting systems is discussed in \textsection{\ref{s:mvss}}.)
The result is $a_1 = -0.08 \pm 0.02$, with a $p$-value of $6\times10^{-5}$. The small $p$-value indicates that there is a statistically significant linear relation between the variability amplitude and the orbital period. Figure~\ref{f:lmfit} shows the data and the best-fitting line.

To better visualize this correlation, Figure~\ref{f:lmfit} also shows the results of dividing the sample into 10 bins, with each bin having roughly the same number of KOIs, and plotting the median $R_{\rm var}$ of the stars within each bin. The uncertainty in each bin's median is roughly $0.026$, estimated as the mean absolute deviation divided by the square root of the number of points in the bin.  Figure~\ref{f:lmfit} also shows the running median, obtained by applying a median filter with a width of 50 data points.  The uncertainty in the running median is $\approx$$0.033$. It appears that the photometric variability decreases significantly with orbital period, particularly for periods longer than about 30~days. Assuming that $R_{\rm var}$ is a proxy for stellar obliquity, as argued by \citet{Mazeh15}, this implies that longer-period planetary orbits tend to be more misaligned with the stellar spin.

The amplitude of photometric variability $R_{\rm var}$ is known to depend on the star's rotation period $P_{\rm rot}$. It is important
to know whether the correlation between $R_{\rm var}$ and $P_{\rm orb}$ could be a side effect of some kind of relationship between $P_{\rm rot}$ and $P_{\rm orb}$.
For this reason we perform another linear regression:
\begin{equation}
P_{\rm rot} = b_0 + b_1 \log{P_{\rm orb}}+ \epsilon .
\end{equation}
The $p$-value of $b_1$ is $0.44$, giving no evidence for a statistically significant linear relation between the stellar rotation period and the planetary orbital period. Thus, the dependence of the photometric variability on planetary orbital period does not seem to be a consequence of an underlying relationship between the stellar rotation period and the photometric variability.

We also computed the Pearson product-moment correlation coefficient between the photometric variability and the orbital period.  The result is $\rho = -0.14$ with the same $p$-value ($6\times10^{-5}$) as the linear regression model.  This also suggests that there is a statistically significant correlation, although the inherent scatter in the photometric variability prevents $\rho$ from approaching the idealized value of $-1$. Between the stellar rotational period and the orbital period, the Pearson correlation coefficient is 0.027, and its $p$-value is 0.44.

The applicability of the linear regression model and the Pearson correlation coefficient depends on the assumption of a linear relation. To avoid this assumption, we calculate Kendall's $\tau$ coefficient and the Spearman rank correlation.  The Kendall's $\tau$ coefficient between photometric variability and the orbital period is $-0.074$, with $p= 0.0017$, and the Spearman rank correlation is $-0.11$, with $p= 0.0018$. Thus, these tests also suggest that there is a statistically significant correlation between the photometric variability of the star and the orbital period of the planet. On the other hand, the coefficients between the rotational period of the star and the orbital period show no statistically significant dependence between the stellar rotational period and the planetary orbital period. Table~\ref{t:pvalues} summarizes
these results.

\begin{table}
\caption{Correlation between photometric variability, stellar rotation period and the orbital period. }
\label{t:pvalues}
\centering
\begin{tabular}{ |l|l|l| }
    \hline
       & $\log{R_{\rm var}}$ \& $\log{P_{\rm orb}}$  & $P_{\rm rot}$ \& $\log{P_{orb}}$   \\ \hline
Pearson's $\rho$   & -0.14 &  0.027 \\ 
p-value & $6.03\times10^{-5}$ & 0.44  \\ \hline
Kendall's $\tau$ & -0.074 & 0.0084 \\ 
p-value & 0.0017 & 0.72 \\ \hline
Spearman's $\rho$ & -0.11 & 0.013 \\
p-value & 0.0017 & 0.71 \\
    \hline
\end{tabular}
\end{table}

\subsubsection{Binning Results}
\label{s:binning}

To further explore the correlation between variability and orbital period, we bin the data according to orbital period.  Then we compare the photometric variability for the KOIs in each period bin with that of the non-KOIs.

\begin{figure}
\includegraphics[width=3.2in]{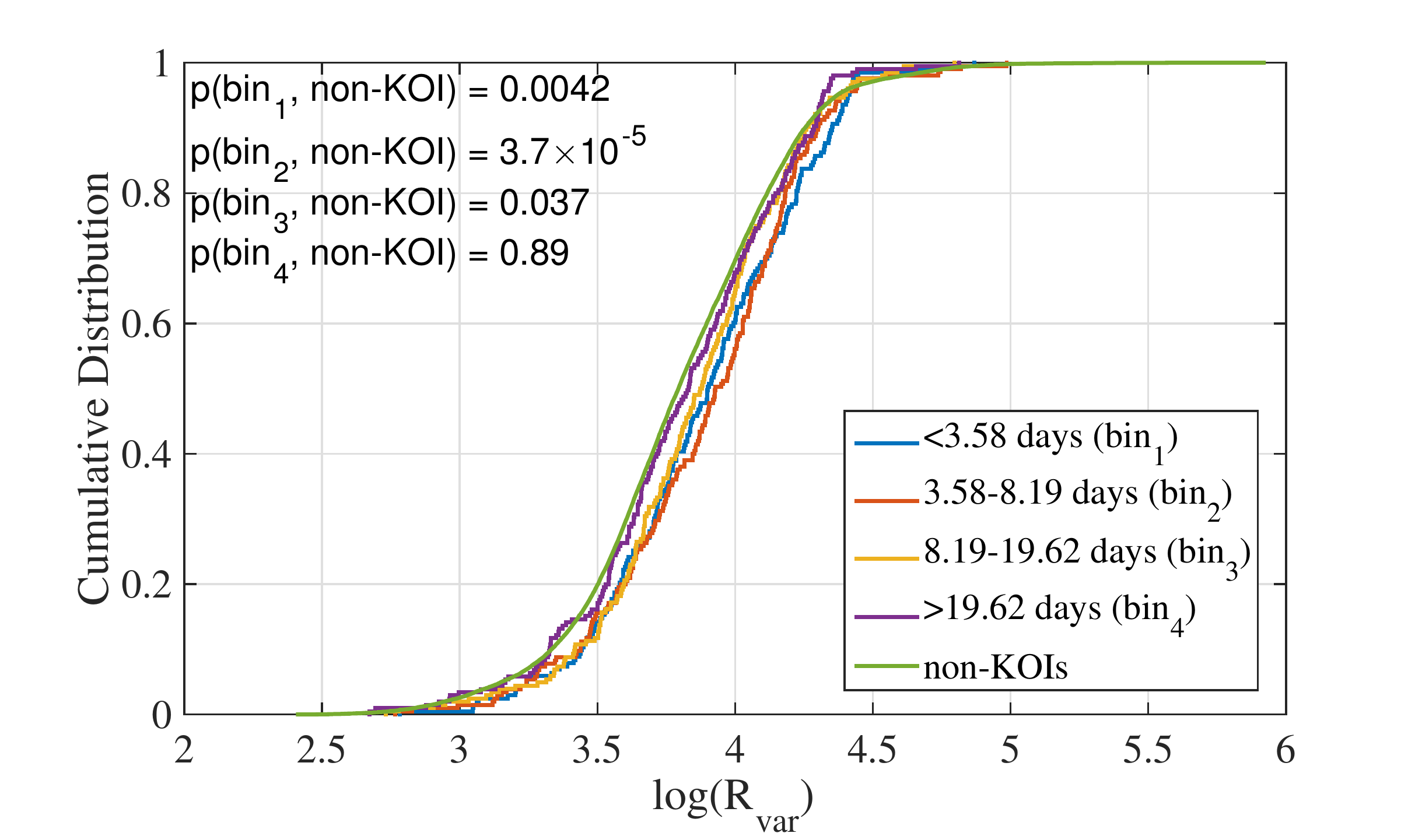} \\
\includegraphics[width=3.2in]{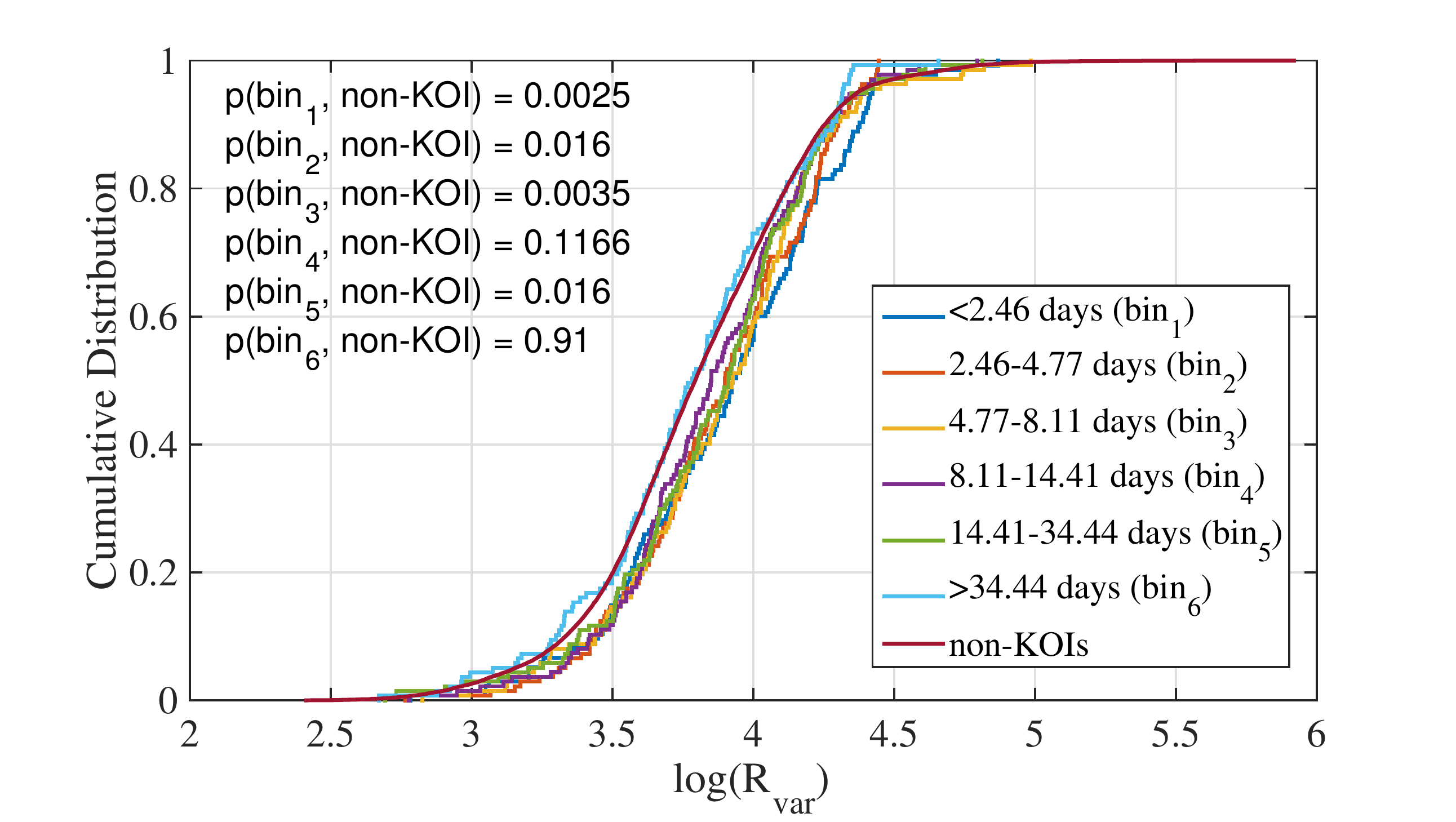}
\caption{Cumulative distribution of the photometric variability for KOIs with different planetary orbital periods and for non-KOIs. We divide the KOIs into four roughly equally sized bins in the upper panel and six bins in the lower panel. It shows that the cumulative distribution of the last bin is similar to that of the non-KOIs, indicating that the spin-orbit misalignment for the last bin is nearly random.}
\label{f:4bin}
\end{figure}

We divide the sample into four bins, with $\approx$200 stars in each bin. The ranges of orbital periods in each bin are $<3.58$ days; $3.58$--$8.19$ days, $8.19$--$19.62$ days, and $>$19.62 days. The median $R_{\rm var}$ values of the stars in each bin are $3.8997\pm0.0192$, $3.9281\pm0.0169$, $3.8760\pm0.0161$, $3.8299\pm0.0181$.  The top panel of Figure \ref{f:4bin} shows the cumulative distribution of $R_{\rm var}$ for the stars in each bin.  We perform KS tests to compare the $R_{\rm var}$ distributions of the KOIs in each bin, and the non-KOIs. In order of increasing orbital period, the $p$-values are $0.0042$, $3.7\times10^{-5}$, $0.0367$, and $0.8928$. Thus, the $p$-value increases for periods longer than $\sim$8 days, and for periods $\gtrsim$20 days there is no evidence for any difference in the $R_{\rm var}$ distribution between the KOIs and non-KOIs.  Again, assuming that enhanced $R_{\rm var}$ is a sign of low obliquity, the evidence for low obliquity is only manifest for periods shorter than a few tens of days.

Next we perform KS tests to compare the $R_{\rm var}$ distribution of the KOIs in the shortest-period bin ($<3.58$ days), and the KOIs in the three longer-period bins. The $p$-values are $0.7598$, $0.5219$, and $0.2029$ for progressively longer-period bins.  The $p$-values are seen to decrease with increasing separation in orbital period, although none of them are low enough for the differences to be considered statistically significant.

To check on the sensitivity of these results to the details of binning, we perform the same tests with 6 period bins, each with about 135 stars.  The ranges of orbital periods in the bins are $<$2.46, 2.46--4.77, 4.77--8.11, 8.11--14.41, 14.41--34.43, and $>$34.43 days. The median $R_{\rm var}$ values of the KOIs in each bin are (in order of increasing period) $3.928\pm0.023$, $3.900\pm0.022$, $3.927\pm0.019$, $3.850\pm0.020$, $3.912\pm0.021$, $3.783\pm0.020$.  The bottom panel of Figure \ref{f:4bin} shows the cumulative distributions of $R_{\rm var}$.  The KS tests comparing the KOIs and non-KOIs give $p$-values of $0.0025$, $0.0160$, $0.0035$, $0.1166$, $0.0163$, and $0.9139$.  The KS test comparing the shortest-period KOIs with the KOIs in the 5 longer-period bins give $p=0.7218$, $0.4969$, $0.3310$, $0.3893$ and $0.0261$.  Table \ref{t:4bin} summarizes these results. Apparently, KOIs with periods longer than $\approx$30 days show similar variability to the non-KOIs, and lower variability than the shortest-period KOIs.

\begin{figure}
\centering
\includegraphics[width=3.2in]{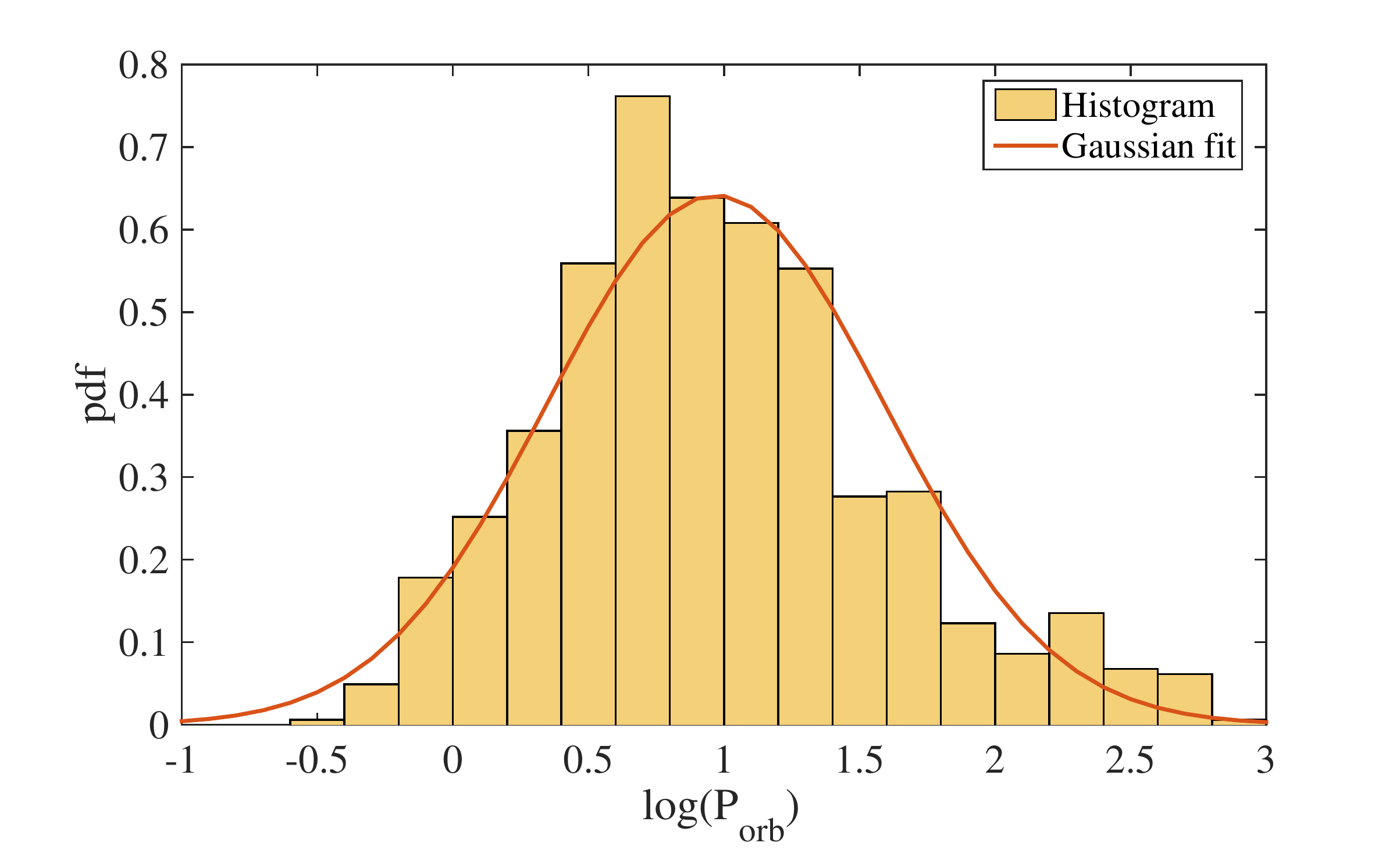}
\caption{Observed period distribution of the KOIs, along with a lognormal fit.
The fit was used to determine the period boundaries of the bins
for which the results are given in Table~\ref{t:4bin_2}.}
\label{f:Gauss}
\end{figure}

We also try an alternative binning method in which the period boundaries are determined by fitting a lognormal function to the distribution of orbital periods (Figure~\ref{f:Gauss}), rather than by simply sorting the KOIs in order of increasing orbital period. Table~\ref{t:4bin_2} gives the results, which are consistent with the preceding results.  Specifically, the KOIs with periods $\gtrsim$30~days exhibit similar variability to the non-KOIs and greater variability than the shortest-period KOIs.

\begin{table*}
\caption{Binning results}
\label{t:4bin}
\centering
\begin{tabular}{|l|l|l|l|l|}
\hline
       & $\rm{bin}_1$  & $\rm{bin}_2$   & $\rm{bin}_3$ & $\rm{bin}_4$\\ \hline
period range  (days) & $< 3.58$  & 3.58-8.19 & 8.19-19.62 & $> 19.62$ \\ \hline
KS p-value w. non-KOIs  & 0.0042 &  $ 3.7\times10^{-5}$ & 0.037 & 0.89 \\ 
KS p-value w. $\rm{bin}_1$  & & 0.76 & 0.52  & 0.20 \\ \hline
WRS p-value w/ non-KOIs & 0.00022 & $3.8\times10^{-5}$ & 0.011 & 0.40 \\ 
WRS p-value w. $\rm{bin}_1$  &  & 0.83  &  0.33  &  0.036 \\ \hline
median  &  $3.90\pm0.02$ & $3.93\pm0.02$ & $3.87\pm0.02$ & $3.83\pm0.02$ \\ \hline 
\end{tabular}
\begin{tabular}{ |l|l|l|l|l|l|l| }
\hline
       & $\rm{bin}_1$  & $\rm{bin}_2$   & $\rm{bin}_3$ & $\rm{bin}_4$ & $\rm{bin}_5$ & $\rm{bin}_6$ \\ \hline
period range  (days) & $<2.46$ & 2.46-4.77 & 4.77-8.11 & 8.11-14.41 & 14.41-34.44  & $> 34.44$ \\ \hline
KS p-value w/ non-KOIs  & 0.0025  &  0.016  &  0.0035  &  0.12  &  0.016  &  0.91 \\
KS p-value w. $\rm{bin}_1$ & &0.72 &   0.50  &  0.33  &  0.39  &  0.026 \\ \hline
WRS p-value w/ non-KOIs & 0.00057 &  0.0019  &  0.0022  &  0.0510  &  0.0179 & 0.74 \\
WRS p-value w. $\rm{bin}_1$ & & 0.63  &  0.59  &  0.21  &  0.32 & 0.0056 \\ \hline
median  &  $3.93 \pm0.02$ & $3.90 \pm 0.02$ & $3.93 \pm 0.03$ & $3.85 \pm 0.02$ &  $ 3.91 \pm 0.02$ &    $3.78 \pm 0.02$ \\ \hline 
\end{tabular}
\end{table*}

\begin{table*}
\caption{Binning results using fitted distribution as shown in Figure \ref{f:Gauss}.}
\label{t:4bin_2}
\centering
\begin{tabular}{|l|l|l|l|l|}
\hline
       & $\rm{bin}_1$  & $\rm{bin}_2$   & $\rm{bin}_3$ & $\rm{bin}_4$\\ \hline
period range  (days) & $< 3.54$  & 3.54-9.31 & 9.31-24.46 & $> 24.46$ \\ \hline
KS p-value w. non-KOIs  & 0.0044 &  $2.4\times10^{-5}$ & 0.032 & 0.8 \\ 
KS p-value w. $\rm{bin}_1$  & & 0.82  &  0.43  &  0.16 \\ \hline
WRS p-value w/ non-KOIs & 0.00021  &  $9.4\times10^{-6}$  &  0.015  &  0.73 \\
WRS p-value w. $\rm{bin}_1$ & & 0.85  &  0.31  &  0.017 \\ \hline
median  &  $3.90\pm0.02$  &  $3.92\pm0.02$  &  $3.88\pm0.02$  &  $3.80\pm0.02$ \\ \hline 
\end{tabular}
\begin{tabular}{ |l|l|l|l|l|l|l| }
\hline
       & $\rm{bin}_1$  & $\rm{bin}_2$   & $\rm{bin}_3$ & $\rm{bin}_4$ & $\rm{bin}_5$ & $\rm{bin}_6$ \\ \hline
period range  (days) & $<2.31$  &  2.31-5.04  &  5.04-9.30  & 9.30-17.16  & 17.16-36.76 & $>36.76$ \\ \hline
KS p-value w. non-KOIs & 0.0014  &  0.0099  &  0.0066  &  0.22  &  0.030  &  0.93 \\
KS p-value w. $\rm{bin}_1$ & & 0.73  &  0.69  &  0.083  &  0.40  &  0.0096 \\ \hline
WRS p-value w/ non-KOIs  & 0.0002  &  0.0014  &  0.0035  &  0.12  &  0.014  &  0.74 \\ 
WRS p-value w. $\rm{bin}_1$ & & 0.41  &  0.33  &  0.062  &  0.39  &  0.0032 \\ \hline
median  &  $3.95\pm0.02$  &  $3.90\pm0.02$  &  $3.89\pm0.02$  &  $3.85\pm0.02$  &  $3.92\pm0.03    $  &  $3.77\pm0.02$ \\ \hline 
\end{tabular}
\end{table*}

In addition, we calculate the Wilcoxon rank sum (WRS) $p$-values, which differs from the KS test by specifically investigating whether one sample tends to have larger values than the other, rather than whether the cumulative distributions are different in any way. The results (also given in Tables~\ref{t:4bin} and \ref{t:4bin_2}) are consistent with the KS tests.

\citet{Mazeh15} did not perform the preceding tests. Instead, they divided the KOIs into two bins ($1-5$ days and $5-50$ days), and calculated the relevant KS $p$-values. We replicate these tests.  In comparing the KOIs with the non-KOIs, we find $p=0.0012$ for the short-period bin and and $1.3\times10^{-4}$ for the long period bin. In comparing the short-period and long-period KOIs, we find $p=0.22$. Thus, we confirm that the two-bin results of \citet{Mazeh15} do not identify any significant period dependence. We have uncovered such a dependence by using more narrowly divided samples in period, and by performing correlation tests on the entire sample.

\subsubsection{Linear Relation vs.\ Step Function}
\label{s:LR}

In principle, the particular form of the period-dependence of the photometric variability (whether linear, nonlinear, or some other functional form) might be a revealing clue about the mechanism for spin-orbit alignment and misalignment.  For instance, tidal dissipation rates are expected to be very strong functions of the star-planet distance, and might therefore produce a sharp drop-off in photometric variability as a function of period. Here, we examine whether the period-dependence of the photometric variability is better fitted by a linear relation, or by a step function (representing an abrupt decrease of photometric variability at a certain orbital period).

\begin{figure}
\includegraphics[width=3.2in]{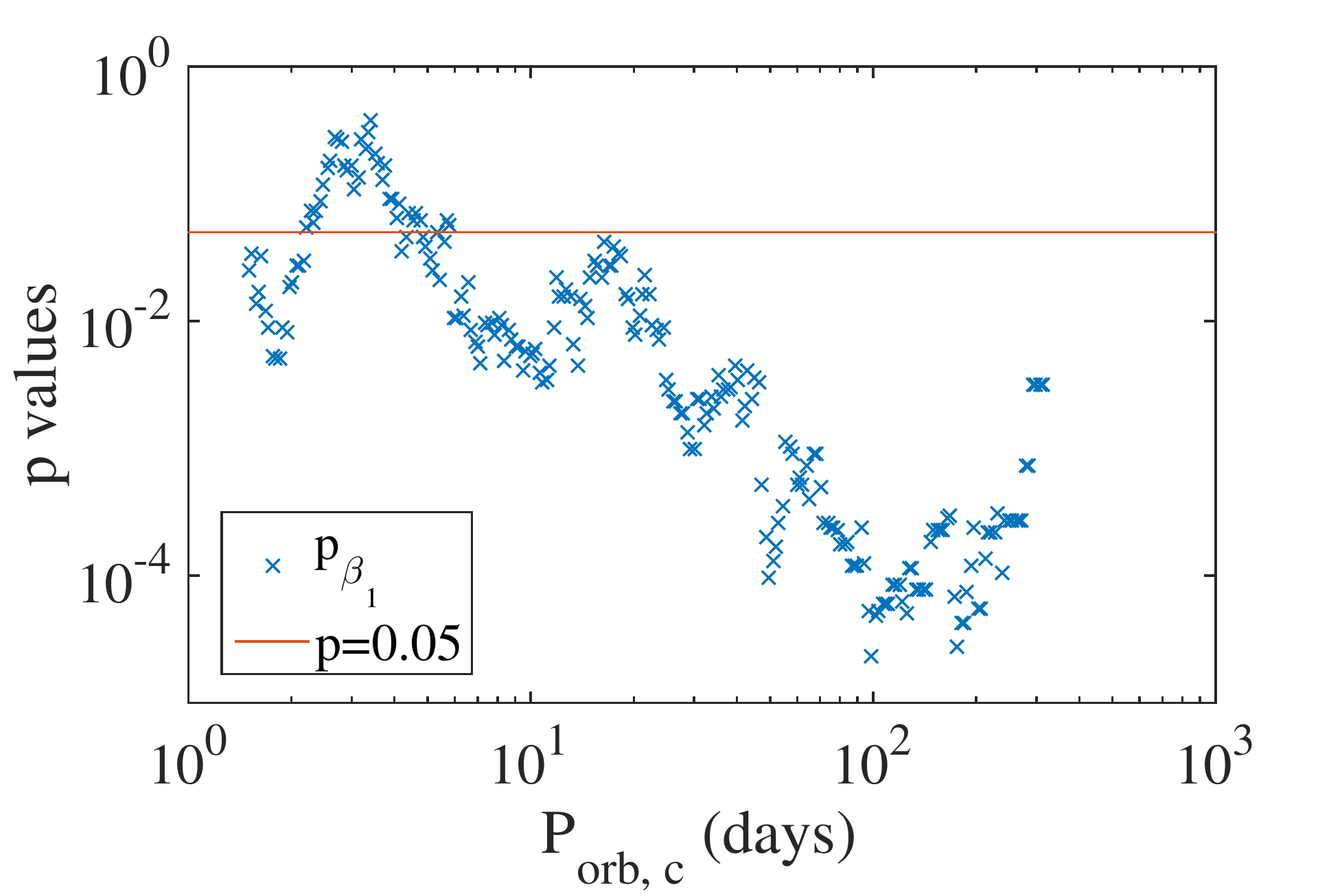} \\
\includegraphics[width=3.2in]{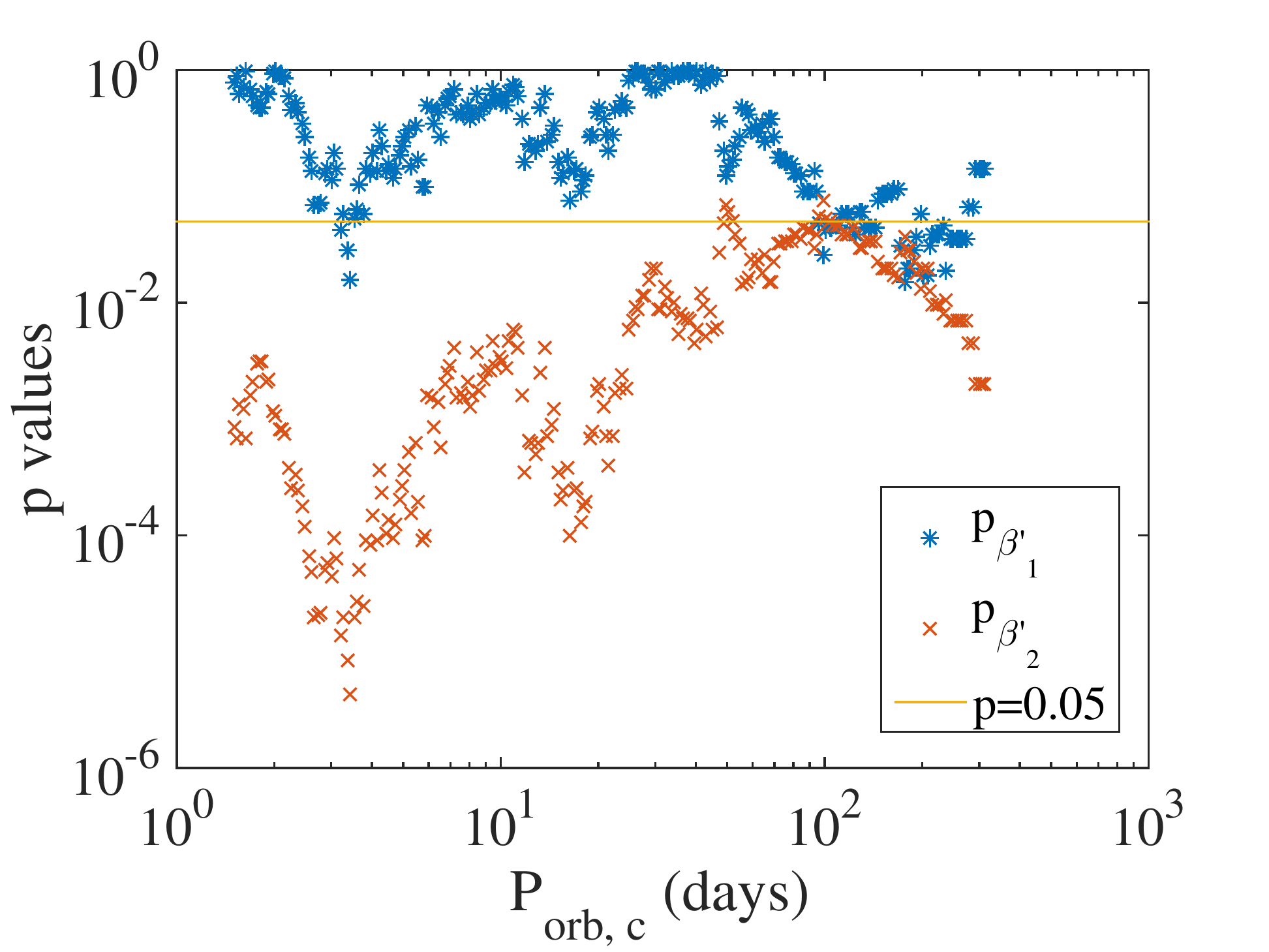} 
\caption{{\it Top.}---The $p$-value for $\beta_1$, as a function of the critical
period $P_{\rm orb, c}$, based on the step-function model of photometric variability
(Eqn.~\ref{e:ind}).
{\it Bottom.}---The $p$-values for $\beta_1'$ and $\beta_2'$, as a function
of the critical period $P_{\rm orb, c}$, based on the multiple linear
regression of the photometric variability including both a linear dependence
and a step-function dependence on orbital period
(Eqn.~\ref{e:mlg}).
Except for $P_{\rm orb, c} \gtrsim 100$ days, the $p$-value for $\beta'_2$ is smaller than 0.05 and the $p$-values for $\beta'_1$ is larger than 0.05, indicating that the linear model
is statistically preferred over the step-function model.}
\label{f:step}
\end{figure}

To test the applicability of the step function, we perform a linear regression
of the photometric variability data with the function
\begin{equation}
\log{R_{\rm var}} = \beta_0 + \beta_1 I_{<P_{\rm orb, c}}(P_{\rm orb}) + \epsilon \,
\end{equation}
where the indicator function $I_{<P_{\rm orb, c}}(P_{\rm orb})$ is defined as
\begin{equation}
I_{<P_{\rm orb, c}}(P_{\rm orb}) = 
\begin{cases}
    1,		& \text{if } P_{\rm orb} < P_{\rm orb, c}\\
    0,		& \text{otherwise} .
\end{cases}
\label{e:ind}
\end{equation}
The $p$-value for $\beta_1$ is plotted in the upper panel of Figure~\ref{f:step}, as a function
of the critical period $P_{\rm orb, c}$. The $p$-value is less than 0.05 everywhere outside of the relatively narrow range $P_{\rm orb, c} =2$-5~days.

To compare the linear model (with respect to $\log{P_{\rm orb}}$) with the step-function model,
we perform a multiple linear regression,
\begin{equation}
\log{R_{\rm var}} = \beta'_0 + \beta'_1 I_{<P_{\rm orb, c}}(P_{\rm orb}) + \beta'_2 \log{P_{\rm orb}} + \epsilon .
\label{e:mlg}
\end{equation}
The $p$-values for $\beta'_1$ and $\beta'_2$ are shown in Figure \ref{f:step}, as a function
of the critical period. Whenever the $p$-value of $\beta'_2$ is less than about 0.05,
the test suggests there is a statistically significant linear relation between
$\log{R_{\rm var}}$ and $\log{P_{\rm orb}}$ even after taking into account any step-function dependence.
On the other hand, when the $p$-value of $\beta'_1$ is below 0.05, the test
suggests that there is a statistically significant step-function dependence of $\log{R_{\rm var}}$ on $P_{\rm orb}$, after taking into account any linear dependence.
The results show that in general (except for $P_{\rm orb, c} \gtrsim 100$ days), $p_{\beta'_2} < 0.05 < p_{\beta'_1}$. This implies that the linear model is preferred over the step-function model,
except when $P_{\rm orb, c} \gtrsim 100$ days.  

As mentioned above, tidal effects are expected to be very strong functions of orbital distance, and to be negligible beyond about 10~days. Quantitatively, then, it may prove difficult for tidal effects to explain the preference for a linear period-dependence over a step-function dependence, or the seemingly abrupt decrease of photometric variability for periods $\gtrsim 100$ days. The data seem to be pointing toward a mechanism that varies more continuously with period, out to $\approx$100 days.  It is also possible that there are several mechanisms affecting spin-orbit alignment, with the net effect producing the dependence on orbital period.

\subsubsection{Selection Effects}
\label{s:se}

Next we consider the possibility that the correlation between $R_{\rm var}$ and $P_{\rm orb}$ is purely a consequence of selection effects.  The difficulty of detecting transiting planets increases with orbital period, as transits become less frequent.  Photometric variability is also a source of noise that potentially interferes with transit detection.  Therefore, it is possible that the long-period KOIs have systematically lower photometric variability than shorter-period KOIs because transits are easier to detect around low-variability stars.  To investigate this possibility we employ a test similar to the one \citet{Mazeh15} performed to check on whether selection effects are responsible for the reduced $R_{\rm var}$ of hot KOIs relative to non-KOIs.

We denote by $\cal P$ the sample of KOIs with $P_{\rm orb} >30$~days, and we denote by $\cal R$ the sample of KOIs with $P<30$~days. We have seen that $\cal P$ and $\cal R$ have differing distributions of photometric variability, and we wish to know if selection effects are wholly responsible. To this end we create simulated samples ${\cal R}_i$ of long-period KOIs for which the photometric variability distribution differs from that of $\cal R$ entirely due to selection effects, through the procedure described below. We then compare the median $R_{\rm var}$ of the simulated KOIs with the median $R_{\rm var}$ of the actual long-period KOIs. These will be indistinguishable, if selection effects are wholly responsible for the differences in $R_{\rm var}$ between $\cal P$ and $\cal R$.

To construct the simulated sample ${\cal R}_i$, we associate with each planet drawn from $\cal P$ a randomly-selected star from $\cal R$ that has favorable enough properties for the planet to have been detected by {\it Kepler}.  In this way, the stars within ${\cal R}_i$ have a distribution of $R_{\rm var}$ that is purely affected by selection effects, and not by any geometrical effects. To decide whether a particular star-planet combination is detectable, we calculate the signal-to-noise ratio, by looking up the signal-to-noise ratio for the actual system from $\cal P$, and scaling it according to
\begin{equation}
S/N \propto \frac{1}{\bar{\sigma}_{\rm CDPP} R_\star^{3/2}} ,
\label{eqn:s2n}
\end{equation}
using the values of $\sigma_{\rm CDPP}$ and $R_\star$ of the randomly-selected star from $\cal R$.
Here, $\sigma_{\rm CDPP}$ is the RMS of the combined differential photometric precision on a 3-hour timescale, and $R_\star$ is the stellar radius. For each star, we use the median of the quarterly $\sigma_{\rm CDPP}$ values. We deem the system to be detectable for $S/N > 10$. (We also experimented with somewhat lower and higher thresholds to ensure that none of the subsequent results depend sensitively on this choice.)

\begin{figure}
\includegraphics[width=3.2in]{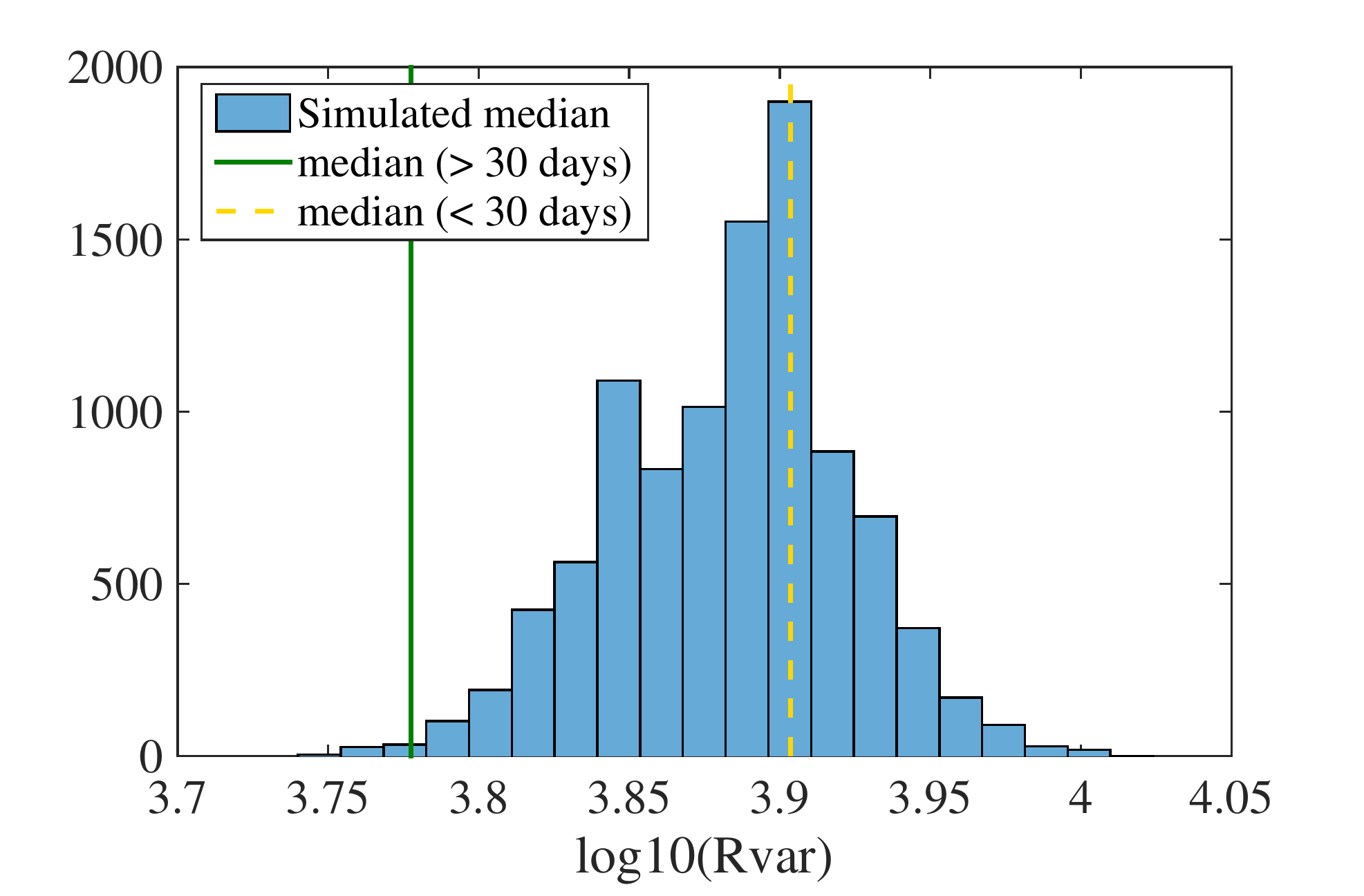}
\caption{Demonstration that selection effects are not wholly responsible for the lower $R_{\rm var}$ of long-period KOIs.  The yellow dashed line shows the median $R_{\rm var}$ of short-period KOIs ($<$30 days).  The green line shows the median $R_{\rm var}$ of long-period KOIs ($>$30~days).  The histogram is the distribution of the median $R_{\rm var}$ of simulated samples of long-period KOIs in which selection effects are entirely responsible for any systematic differences in variability. The spread in this distribution is not large enough for selection effects to be a plausible explanation of the entire difference between the short and long-period KOIs.}
\label{f:se}
\end{figure}

We construct $10^4$ different random realizations of ${\cal R}_i$. Figure~\ref{f:se} shows the resulting distribution of the median values of $R_{\rm var}$. Only $51$ out of the $10^4$ simulated samples have a median $R_{\rm var}$ that is lower than the median $R_{\rm var}$ of the real long-period KOIs. Thus, it is unlikely ($\sim$0.5\%) that the low variability associated with the long-period KOIs is purely a consequence of selection effects.

\subsubsection{Geometrical Interpretation}
\label{s:toy}

Having ruled out selection effects as the entire explanation for the lower variability of longer-period KOIs, we now examine the expected amplitude of the geometrical effect in slightly more detail than \citet{Mazeh15}. Those authors supposed that the amplitude of photometric variability of a sample of stars scales as the mean value of $\sin i_\star$, where $i_\star$ is the inclination of the stellar spin axis with respect to the line of sight. This scaling seems reasonable for an individual star, and is supported by simulations of spotted stars of varying inclinations by \citet{JacksonJeffries13}. Under this assumption, the mean\footnote{In this section we examine the mean, rather than the median, for computational simplicity, and because the distinction is not significant in the subsequent discussion.} $R_{\rm var}$ of a population of randomly-oriented stars should be lower by a factor of $\pi/4$ (the mean value of $\sin i$ for random points on a sphere) than a population of stars with $i_\star=90^\circ$.  This corresponds to a difference in $\log R_{\rm var}$ of about 0.1 dex, in agreement with the observed difference between the short-period and long-period KOIs.

Although these limiting cases of $i_\star=90^\circ$ and random obliquities are easy to evaluate, it was not initially clear to us how the {\it spread} in obliquities of a population of stars is related to the mean value of $\sin i_\star$. This is relevant to the question of whether the long-period KOIs must be very nearly random in orientation or whether a small but nonzero spread in obliquities is sufficient to account for the variability statistics. 

We assume as above that the photometric variability $R_{\rm var}$ is proportional to $\sin{i_\star}$. We write $i_*$ as a function of $\theta$ and $\psi$, where $\theta$ is the obliquity, and $\phi$ is the longitude of the stellar spin axis with respect to the planetary orbital plane:
 \begin{equation}
 \cos{i_\star} = \sin{\theta}\cos{\phi} .
 \end{equation}
 
We assume that $\phi$ is uniformly distributed:
\begin{equation}
f_\phi (\phi) = \frac{1}{2\pi}.
\end{equation}
Following \citet{FabryckyWinn09} and \citet{Morton14} we assume the obliquity $\theta$ obeys a Fisher distribution (also known as
a $p=3$ von Mises-Fisher distribution):
\begin{equation}
 f_\theta(\theta | \kappa) = \frac{\kappa}{2\sinh{\kappa}} e^{\kappa \cos{\theta}} \sin{\theta}, \label{eqn:fish}
\end{equation}
where the concentration parameter $\kappa$ controls the spread in obliquity.
For large $\kappa$, $f_\theta$ approaches Rayleigh distribution with width $\sigma \to \kappa^{-1/2}$, and when $\kappa \to 0$, the distribution is isotropic.  

\begin{figure}
\includegraphics[width=3.2in]{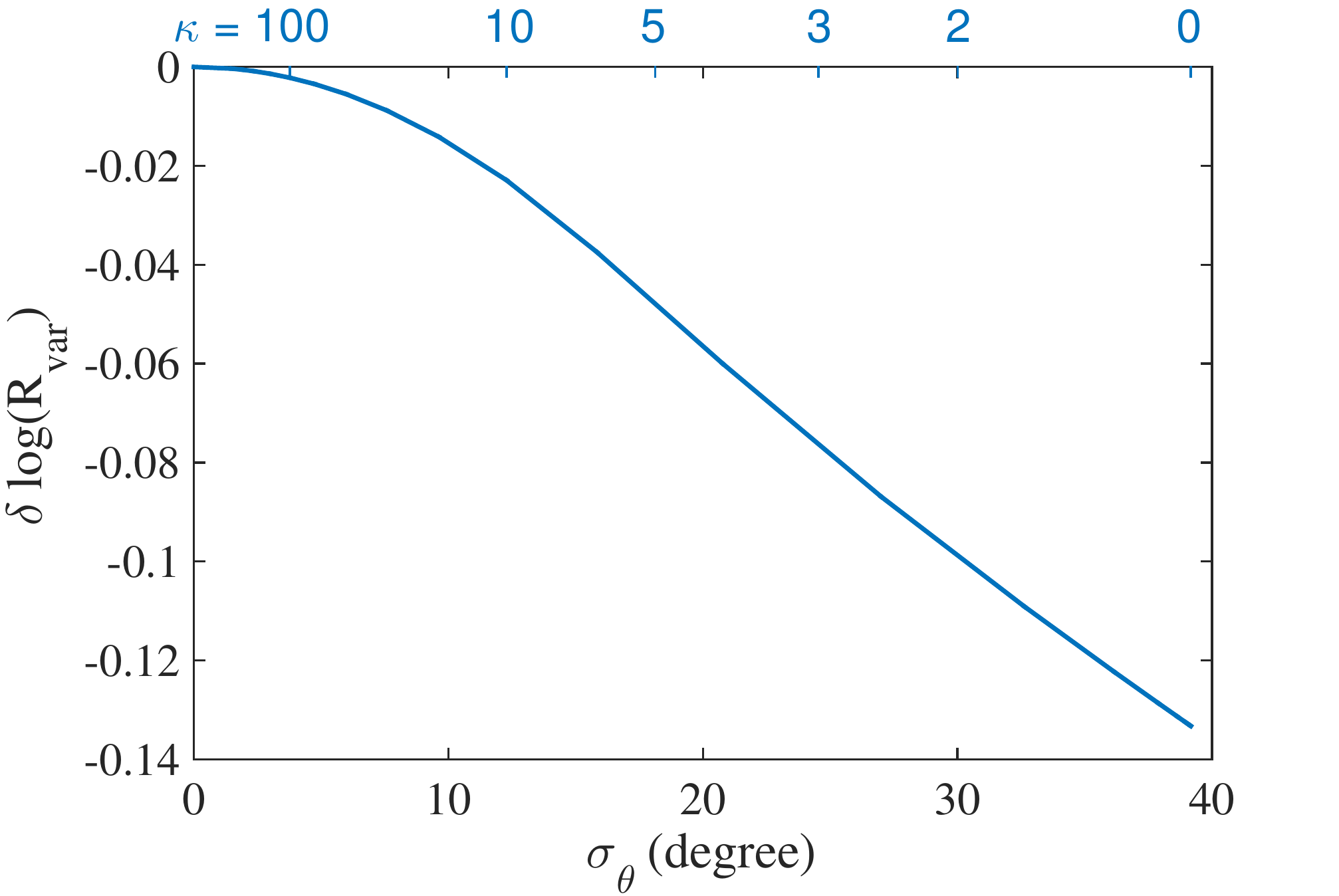}
\caption{Expected relationship between photometric variability and the spread in obliquities of a population of stars. Plotted is $\delta \log{R_{\rm var}}$ versus the concentration parameter $\kappa$ of the Fisher distribution (top axis, Eqn.~\ref{eqn:fish}) as well as the standard deviation $\sigma_\theta$ of the obliquity (lower axis).}
\label{f:fisher}
\end{figure}

Then, we calculate the expectation value of $\sin i_\star$ as a function of $\kappa$, as well as the expected difference in $\log{R_{\rm var}}$ compared to the case of $\kappa\rightarrow \infty$ (zero obliquity).
Specifically, 
\begin{eqnarray}
& & \langle \sin{i_*}_\kappa \rangle \\ \nonumber
&=&\int_0^\pi\int_0^{2\pi} \sqrt{1-\sin^2{\theta}\cos^2{\phi}}f_\theta(\theta | \kappa) f_\phi(\phi)~d\phi~d\theta \\
& &\delta \log{R_{\rm var}}_\kappa = \langle \log{\sin{i_*}_\kappa }\rangle  \, .
\end{eqnarray}
Figure~\ref{f:fisher} shows the results for $\delta \log{R_{\rm var}}$ as a function of $\kappa$ (on the upper axis) and the standard deviation of $\theta$ (on the lower axis).  As expected, when the obliquity distribution is isotropic ($\kappa=0$) we obtain $\delta \log{R_{\rm var}} \approx -0.13$.

Based on the results of our 4-bin analysis (Figure \ref{f:4bin} and Tables~\ref{t:4bin} and \ref{t:4bin_2}), the observed difference $\delta \log{R_{\rm var}}$ between the shortest-period KOIs ($< 3.58$ days) and longest-period KOIs ($>$19.62~days) is $0.0698\pm0.0373$.  Therefore, if the shortest-period KOIs are assumed to have zero obliquity, and the longest-period KOIs obey a Fisher distribution, then the observations suggest that $\kappa \lesssim 10$ or $\sigma_\theta \gtrsim 15^\circ$ for the longest-period KOIs.  In our 6-bin analysis, we found $\delta \log{R_{\rm var}}=0.1458\pm0.0429$ between the shortest-period KOIs ($<$2.46~days) and the longest-period bin ($>$34.43~days). This suggests that for periods longer than $\approx$30~days, the obliquity distribution is nearly random.  Given that the expected deviations $\delta \log{R_{\rm var}}$ are so small compared with the inherent scatter in the photometric variability, it is difficult to go beyond these relatively crude results. The translation between $\delta \log{R_{\rm var}}$ and $\kappa$ (or $\sigma_\theta$) may be useful when expanded samples of transiting planets are available.

\subsubsection{Multi-transiting vs.\ Single-transiting Systems}
\label{s:mvss}

The preceding investigations did not make any distinction between KOIs with only one detected transiting planet (``singles''), and KOIs with multiple transiting planets (``multis'').  However, there may be inherent differences between these two samples. For instance, based on estimates of projected rotation rate, rotation period, and stellar radius, \citep{Morton14} found tentative evidence that multis have lower obliquities than singles. Here we search for such differences with the photometric technique of \citet{Mazeh15}.  Our sample includes 606 singles and 208 multis (with 524 planets).

We first seek evidence within the sample of multis for any period dependence of the stellar variability. Among the multis, the orbital periods of the innermost planets extend to 86 days. Applying the linear regression model of Eqn.~(\ref{e:linear}) we find $p=0.56$, indicating no statistically significant relation between $R_{\rm var}$ and the period of the innermost planet.  In contrast, the linear model applied to the singles gives $p=5\times10^{-5}$, indicating a significant dependence of $R_{\rm var}$ upon $P_{\rm orb}$.

Then, we compare the variability of 3 subsamples: single-transiting KOIs with $P_{\rm orb}<30$~days (``short-period singles''), single-transiting KOIs with $P_{\rm orb}>30$~days (``long-period singles''), and KOIs with multiple detected transiting planets (``multis'').  When comparing the $R_{\rm var}$ distributions of the short-period singles and the multis, the KS test gives $p=0.82$ and the WRS test gives $p=0.51$, indicating no discernible statistical difference. On the other hand, when comparing the long-period singles and the multis, the KS test gives $p=0.018$ and the WRS test gives $p=0.0079$, pointing to a significant difference. Table \ref{t:multi} summarizes these results, which do not depend critically on the choice of cutoff period; the table also shows the results when the sample is divided at a period of 20 days and 40 days instead of 30 days.

Next we compare the singles and the multis with the non-KOIs. The photometric variability distribution of the short-period singles is significantly different from that of the non-KOIs (KS $p=1.6\times 10^{-7}$, WRS $p=7.7\times 10^{-9}$). The multis also have a different $R_{\rm var}$ distribution from the non-KOIs (KS $p=0.0091$, WRS $p=0.0019$). But these tests do not reveal any difference between the long-period singles and the non-KOIs.  Table~\ref{t:multi} summarizes these results.

\begin{table}
\caption{Single-transiting v.s. multi-transiting planetary systems }
\label{t:multi}
\centering
\begin{tabular}{ |l|l|l| }
    \hline
       & WRS p-value  & KS p-value   \\ \hline
Innermost multi v.s. Single ($<20$ days)  & 0.54 & 0.84 \\ \hline
Innermost multi v.s. Single ($<30$ days)  & 0.51 &  0.82 \\ \hline
Innermost multi v.s. Single ($>30$ days)  & 0.0079 &  0.018 \\ \hline 
Innermost multi v.s. Single ($>40$ days)  & 0.011 & 0.026 \\ \hline
Multi periods v.s. Single periods & 0.41  & 0.078 \\     \hline
Innermost multi v.s. non-KOIs & 0.0019 & 0.0091 \\ \hline
Single ($<30$ days) v.s. non-KOIs & $7.7 \times 10^{-9}$ & $1.6\times 10^{-7}$ \\ \hline
Single ($>30$ days) v.s. non-KOIs & 0.36 & 0.53 \\ \hline
\end{tabular}
\end{table}

\begin{figure}
\includegraphics[width=3.2in]{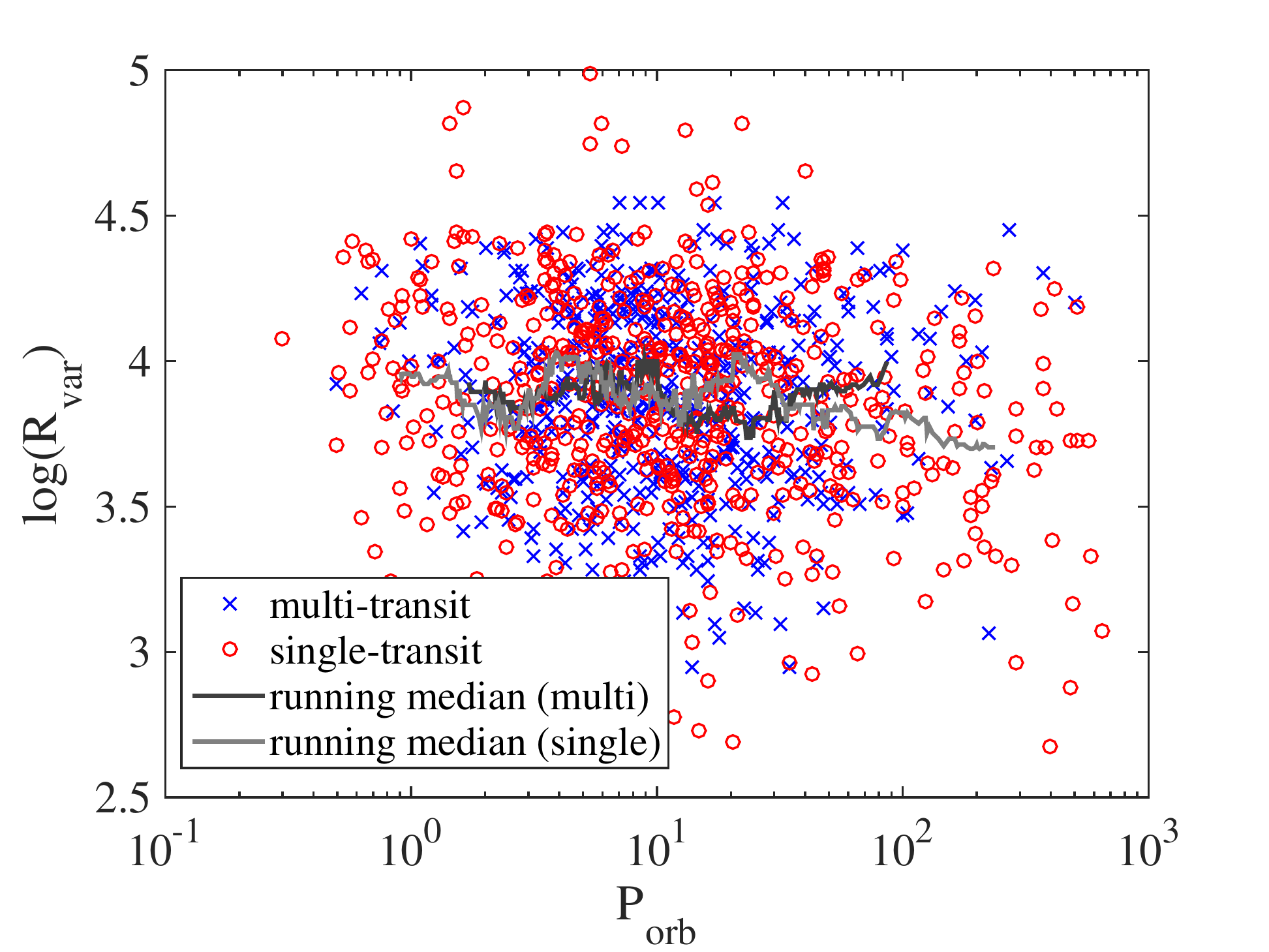}
\caption{Photometric variability $R_{\rm var}$ versus planetary orbital period, for the host
stars of multple detected transiting planets (blue crosses) and
the host stars with only one detected transiting planet (red circles).
The black line represents the running median of the multi-transiting sample and the gray line represents the running median of the single-transiting sample. For both lines, the width of the running median filter is 50 data points and the uncertainty in the median is about 0.033.}
\label{f:lmfit_multi}
\end{figure}

These findings are illustrated in Figure \ref{f:lmfit_multi}, which shows $R_{\rm var}$ as a function of the orbital period for the multis and the singles separately. For short orbital periods, the variablility of singles and multis are indistinguishable. For longer orbital periods ($\gtrsim$30~days) the multis seem to show enhanced variability relative to the singles. Interpreting these findings in terms of geometry, it seems that the multis and short-period singles both have host stars that tend to have low obliquities. Once the orbital period exceeds a few tens of days, the singles begin to have a broader obliquity distribution, while the multis retain lower obliquities.

The dependence of the misalignment on the orbital period for the singles, and the alignment of the multis independently of period, may have implications for the proposed theories in which spin-orbit misalignment is a consequence of mechanisms that tilt a star away from the protoplanetary disk. We have in mind the tilting of the stellar spin axis due to the internal oscillation modes of the star or magnetic field-disk interactions \citep{Rogers12, Lai11, Spalding14}, or the tilt of the protoplanetary disk due to a distant companion or the turbulent stellar formation environment and interactions between stars \citep{Batygin12, Bate10, Thies11, Fielding15}. Naively we would not expect such mechanisms to depend on the ultimate configuration of the planets in the system, in contradiction with our findings in this section. However there are probably ways to avoid this contradiction. For example, a tilted disk may become warped such that its inner region is realigned with the star, and thereby lead to planetary systems that are not coplanar and unlikely to be observed as multi-transiting systems.  (We note, though, that \citet{Lai11} and \citet{Foucart11} have investigated the possibility of inner disk warps, and found that they are difficult to maintain due to viscous stresses and bending waves.) On the other hand, the mechanisms which tilt the star through star-planet dynamics seem very likely to depend sensitively on the planetary orbital configuration and are probably more applicable to the singles; these properties appear more in accordance with our findings.

\section{Tidal Re-alignment}
\label{s:tide}

In the previous section we examined an observational challenge to the notion that star-planet tidal interactions are responsible for enforcing spin-orbit alignment.  The observation of enhanced photometric variability for relatively cool host stars of short-period transiting planets, and the absence of this enhanced variability in systems with longer orbital periods, are perhaps qualitatively compatible with tidal effects. Quantitatively, though, it is difficult to understand how the effects could be important for relatively small planets with orbital periods as long as $\sim$10~days.

In addition, as described in \S~1, the tidal hypothesis presents theoretical difficulties: the same tidal interactions that coplanarize the system should also result in angular momentum being transferred from the orbit to the stellar spin, causing the orbit to shrink and the planet to be destroyed \citep{Winn10}. Recent work on tidal theory suggests that excitation and dissipation of inertial waves play an important role in stellar spin and orbital evolution \citep[e.g.,][]{Ogilvie05, Favier14}. \citet{Lai12} recognized that inertial waves offer a possible means for a planet to realign its host star without being ingested. He identified a component in tidal potential (attributed to inertial waves) with a frequency equal to the stellar spin frequency in the rotating frame of the star. Because it is static in the inertial frame, this component acts to reduce the obliquity without causing orbital decay.

\citet{Rogers13b} called this solution into question, by pointing out that this particular component of the tidal potential does not lead exclusively to prograde configurations.\footnote{\citet{Ogilvie13, Ogilvie14} have also pointed out that there may be no dissipation associated with this tidal component because it is a spin-over mode with a uniform rotation, while also acknowledging that there may be non-trivial dynamics involving the core or the precession of the spin axis of the star that may lead to dissipation. We do not attempt to address this important issue in this paper.} Considering both this component and the equilibrium tide, and requiring the orbital decay timescale to be longer than the obliquity alignment timescale, \citet{Rogers13b} found that initially prograde systems evolve to an obliquity of 0$^\circ$, while initially retrograde systems evolve to either $90^\circ$ or $180^\circ$.  Thus, starting with a uniform distribution of the obliquity, tidal evolution produces retrograde and perpendicular systems half of the time. This is in strong contrast to the observations, in which prograde configurations are by far the most common \citep{Albrecht12}.

Here we show that tidal evolution can lead to spin-orbit re-alignment before rapid orbital decay, even when starting from an initially retrograde configuration taking into account magnetic braking. Our calculations differ from those of \citet{Rogers13b} mainly by considering the evolution of the system over longer timescales. We find that even though the obliquity may stall at $90^\circ$, the obliquity continues to decreases slowly due to the equilibrium tide. Eventually, the equilibrium tide dominates and quickly brings the system into alignment. Depending on the ratio of the rates of tidal dissipation due to inertial waves and the equilibrium tide, it is possible for spin-orbit alignment to be achieved more rapidly than orbital decay. For the case in which the obliquity stalls near $180^\circ$, the equilibrium tide reduces the stellar spin rate and decreases the spin angular momentum of the star.  Eventually the stellar spin angular momentum becomes much smaller than that of the planetary orbit, reducing the timescale for the equilibrium tide to align the star with the orbit.  \citet{Xue14} also found that the combined effects of the equilibrium tide and inertial waves can lead to tidal re-alignment prior to orbital decay, even from an initially retrograde state. Our study differs by including the effects of magnetic braking. We have also delineated the parameter space within which the re-alignment can be achieved before orbital decay.

\begin{figure}
\includegraphics[width=3.2in]{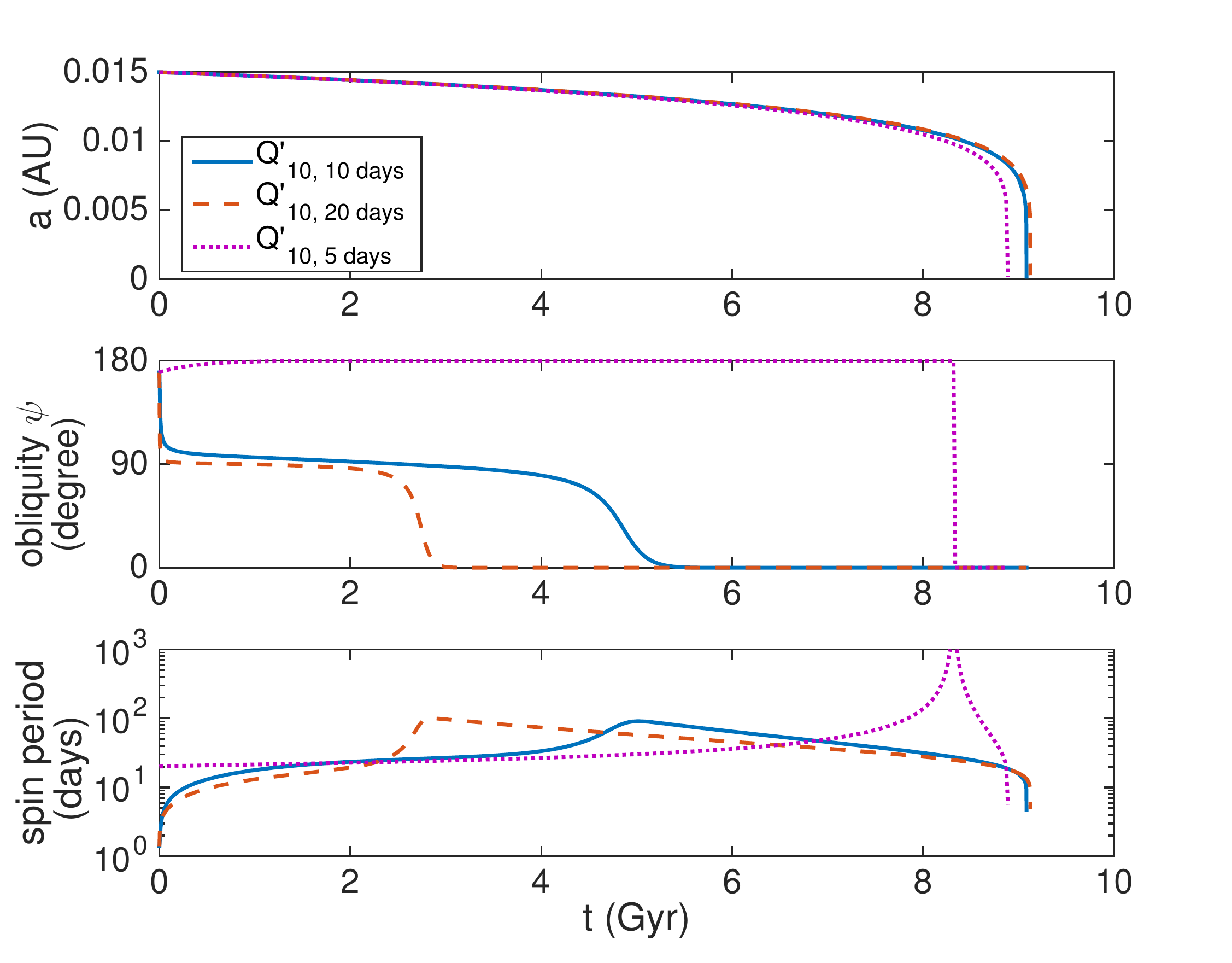}
\caption{The re-alignment of the stellar obliquity, the decay of the orbit and the change of the stellar spin period as a function of time. It shows that the stellar spin axis can be aligned with the planetary orbit and can survive for a long period of time before rapid orbital decay under equilibrium tidal effects, magnetic braking, and the effects produced by the tidal component identified by \citet{Lai12}.
}
\label{f:tide}
\end{figure}

We begin with an illustrative example of the obliquity evolution of an initially retrograde system.  In addition to tidal effects we also take into account the loss of stellar angular momentum due to magnetic braking. We integrate the equations for the evolution of the orbital distance ($a$), stellar spin angular frequency ($\Omega_s$) and obliquity ($\Theta$) that were presented by \citet{Lai12} (Eqn. 57-59),
summarized as
\begin{align}
&\frac{\dot{a}}{a} = -\frac{1}{t_a} \Big(1-\frac{\Omega_s}{\Omega}\cos{\Theta} \Big) ~, 
\label{e:t1}\\
&\dot{\Omega}_s = \dot{\Omega}_{s, e} +  \dot{\Omega}_{s, 10} -  \dot{\Omega}_{s, 10, e} + \dot{\Omega}_{s, m} ~, \\
&\dot{\Theta} = \dot{\Theta}_e +  \dot{\Theta}_{10} -  \dot{\Theta}_{10, e}  ~, 
\label{e:t2}
\end{align}

where 
\begin{align}
&\Big( \frac{\dot{\Omega}_s}{{\Omega_s}} \Big)_e = \frac{1}{t_a} \Big(\frac{L}{2S} \Big) \big[\cos{\Theta}-\Big(\frac{\Omega_s}{2\Omega}\Big)(1+\cos^2{\Theta})\big] ~, \\
&\Big( \frac{\dot{\Omega}_s}{{\Omega_s}} \Big)_{10} = -\frac{1}{t_{s10}}(\sin{\Theta}\cos{\Theta})^2 ~, 
\label{e:o10}\\
&\dot{\Omega}_{s, m} = -\alpha \Omega_s^3 ~, \\
&\dot{\Theta}_e = -\frac{1}{t_a} \Big(\frac{L}{2S}\Big) \sin{\Theta} \Big[1-\Big(\frac{\Omega_s}{2\Omega}\Big)\Big(\cos{\Theta}-\frac{S}{L}\Big)\Big] ~, \\
&\dot{\Theta}_{10} =  -\frac{1}{t_{s10}} \sin{\Theta}\cos^2{\Theta}\Big(\cos{\Theta}+\frac{S}{L}\Big) ~, 
\label{e:t10}\\
&\frac{1}{t_a} = \frac{3k_{2e}}{Q_e}\Big(\frac{M_p}{M_s}\Big)\Big(\frac{R_s}{a}\Big)^5 \Omega ~, \\
&\frac{1}{t_{s10}} = \frac{3k_{2,10}}{4Q_{10}}\Big(\frac{M_p}{M_s}\Big)\Big(\frac{R_s}{a}\Big)^5\Big(\frac{L}{S}\Big)\Omega ~. 
\end{align}
The expressions for $\dot{\Omega}_{s, 10, e}$ and $\dot{\Theta}_{10, e}$ are the same as Eqn.~(\ref{e:o10}) and (\ref{e:t10}), except that $Q_{10}$ and $k_{2,10}$ are replaced by $Q_e$ and $k_{2,e}$. 
The subscript ``$e$'' stands for equilibrium tide.
The subscript ``$10$'' denotes the tidal component identified by \citet{Lai12};
this nomenclature is based on the fact that this component has $m=1$ and $m'=0$ in
the general expression for the the tidal forcing frequency, $\varpi = m' \Omega - m \Omega_s$
(where $\Omega$ is the orbital angular velocity). $L$ stands for the angular momentum of the planetary orbit, and $S$ stands for that of the stellar spin. $M_s$ denotes the mass of the star, $M_p$ denotes the mass of the planet, and $R_s$ denotes the radius of the star.
For this example we set $M_s=M_\odot$, $R_s=R_\odot$, $M_p = M_{\rm Jup}$, and $\alpha = 1.5\times10^{-14}$~yr, which gives a braking timescale of $2\times10^{11}$ yr for the Sun as discussed by \citet{Barker09}.

The studies of energy dissipation in convection zone by \citet{Ogilvie07} show that inertial waves can increase the tidal dissipation rate by up to four orders of magnitude.  For the $m=2$ component a value as small as $Q'= (3Q)/(2k_2) = 10^6$ can be achieved, for stars rotating more rapidly than the Sun.
The tidal dissipation rate has a complicated dependence on the tidal frequency, and varies with
the spin rate as $\Omega_s^{-2}$ for $\Omega_s \ll \omega_d$, where $\omega_d$ is the dynamical frequency of the star \citet{Ogilvie07}. For our illustrative example, we adopt a constant $Q_e=5\times10^7$ and $k_2=0.028$ ($Q'_e = 2.7\times10^{9}$), and we set $Q'_{10} = 10^6 (\Omega_s/\Omega_{\rm s, k~{\rm days}})^{-2}$, where $\Omega_{s, k~{\rm days}}$ is the angular velocity of the stellar spin corresponding
to a period of $k$ days. The decrease of $Q'_{10}$ relative to $Q'_e$ will lead to a shorter re-alignment timescale at $\sim 90$ degree comparing with the orbital decay timescale, allowing a higher likelihood for the aligned configuration to be observed. We explore a large range of possible
values of $Q'_{10}$ and investigate the outcomes.

Figure \ref{f:tide} shows the evolution of the orbital distance, stellar obliquity, and stellar spin period due to tidal interactions and magnetic braking. We set the initial orbital distance to be 0.015 AU, the initial obliquity to be $170^\circ$ and the initial spin period to be 1.4 days.  Because $\Omega_s < \Omega / 2$, inertial wave dissipation is forbidden and orbital decay is prolonged.  The solid blue line represents the case when $Q'_{10}=10^6 (\Omega_s/\Omega_{s, 10})^{-2}$, and the dashed orange line represents the case when $Q'_{10}=10^6 (\Omega_s/\Omega_{s, 20})^{-2}$. For these two cases, the obliquity quickly reaches $90^\circ$ and stalls for 3-5~Gyr before eventually damping to zero.  The orbital distance stays near the initial value before rapidly decaying after 9~Gyr.  Thus, in this case, there is an interval of a few billion years when it is possible to observe spin-orbit alignment before orbital decay.  For smaller values of $Q'_{10}$, this interval is prolonged, and the stellar spin can be re-aligned with the planetary orbit at greater orbital distances.

Both magnetic braking and the $(1,0)$ tidal component act to increase the spin period.  Initially, when the period is short, magnetic braking is the dominant effect.  After the obliquity stalls at $90^\circ$ and begins its more gradual descent to zero, the despinning effect of the $(1,0)$ component also becomes important.  Eventually, as $a$ and $\Omega_s$ shrink, the spin-up due to the equilibrium tide becomes dominant and the spin period decreases.  Thus, the interval between spin-orbit alignment and orbital decay is characterized (at least at first) by a relatively long rotation period.  Interestingly, this is qualitatively consistent with the observation of \citet{McQuillan13} that rapidly-rotating stars seem to be missing short-period planets, relative to their occurrence around slowly-rotating stars (see also \citealt{Teitler14}).

The dotted purple line in Figure \ref{f:tide} shows the case of $Q'_{10}=10^6 (\Omega_s/\Omega_{s, 5})^{-2}$. In this case the obliquity is driven to $180^\circ$ rather than $90^\circ$, because of the smaller value of $S/L$. At $180^\circ$, the $(1,0)$ tidal component does not change the spin rate or the obliquity. The stellar spin frequency is reduced by magnetic braking and the equilibrium tide, the angular momentum of the stellar spin becomes much smaller than that of the planetary orbit, and the timescale to synchronize and align the stellar spin becomes short. As shown in Figure \ref{f:tide}, the stellar spin can be torqued from the retrograde state to the prograde state before rapid orbital decay. An interesting feature of this scenario is that the star's rotation is slowed to a standstill (as indicated in Fig.~8 by the sharp rise in rotation period), before being spun up in the other direction.  Thus, while previous searches for planet-star interactions have been based on the expectation that tidally influenced host stars will rotate more rapidly (see, e.g., \citealt{Pont09}), it is also possible that they will exhibit unusually slow rotation.
  While a broad range of rotation periods is observed for stars of a given age and mass \citep{Angus15}, we are not aware of any evidence for slower rotation among any particular category of exoplanet host stars.

We explore a wider range of parameter space by evolving the obliquity using Eqn.~(\ref{e:t1} -\ref{e:t2}) for different values of the initial stellar spin period and $Q'_{10, 10~{\rm days}} / Q'_e$. We adopt fixed values $a_0=0.015$~AU and $Q'_e = 2.7\times10^9$. Thus in all cases the orbit decays rapidly after $\approx 9$~Gyr.  We set $Q'_{10} = Q'_{10, 10~{\rm days}} (\Omega_s/\Omega_{s, 10})^{-2}$, and we explore initial values of $\Omega_s / \Omega$ ranging from 0.05--0.5. When $\Omega_s > \Omega / 2$, other inertial waves can be excited, leading to more rapid orbital decay. The other parameters are the same as those for the illustrative cases described above. To quantify the timescales of alignment and orbital decay, we record the ratio of the alignment timescale and the orbital decay timescale. We also calculate the spin period of the star during the interval between spin-orbit alignment and orbital decay.

\begin{figure}
\includegraphics[width=3.2in]{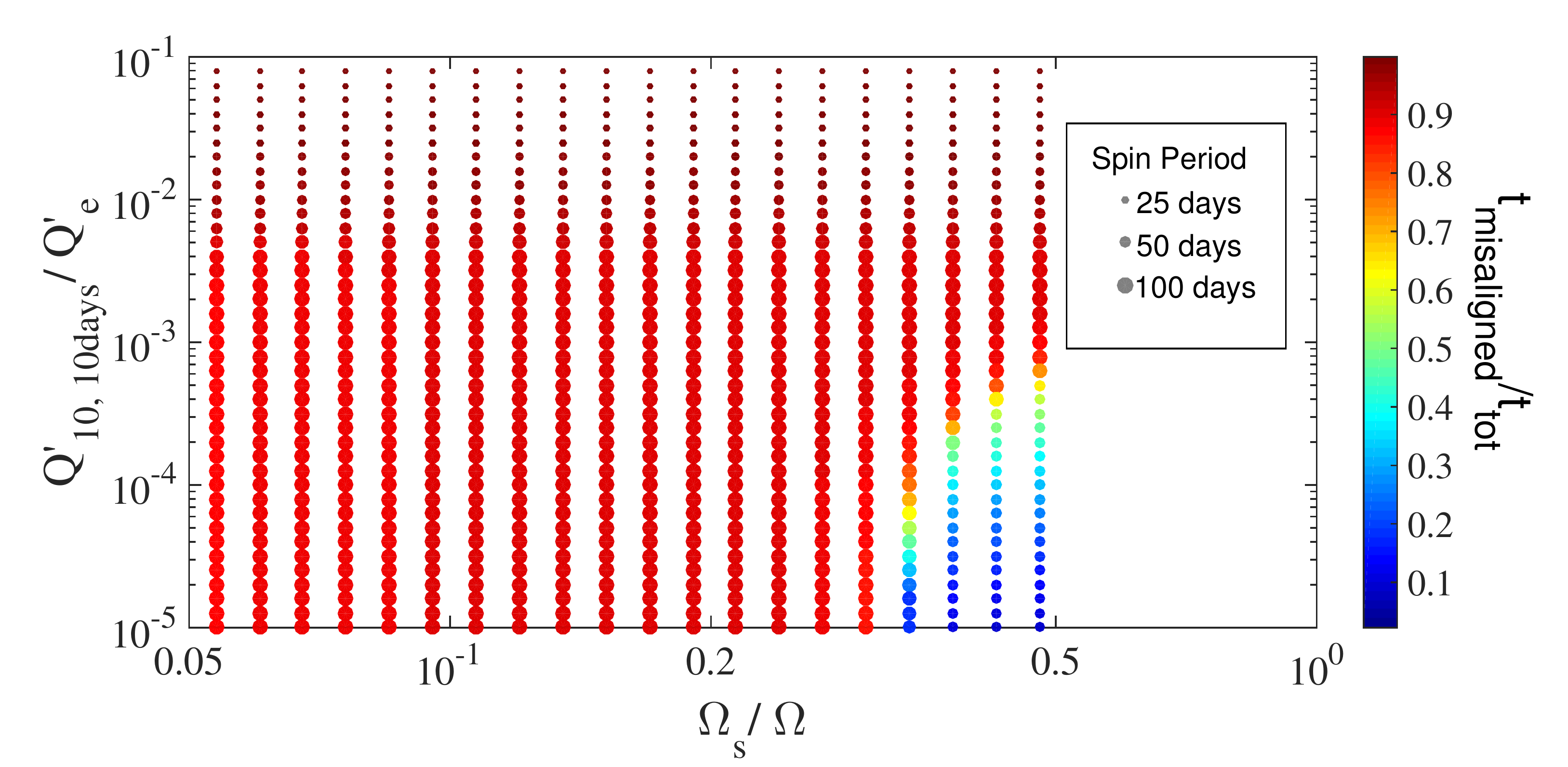} 
\caption{Ratio of spin-orbit alignment timescale and orbital decay timescale (color coded),
as a function of the initial value of $\Omega_s/\Omega$ and $Q'_{10, 10~{\rm days}} / Q'_e$.
Low ratios (blue) lead to a longer time interval during which spin-orbit alignment
can be observed before orbital decay. The size of each circle is proportional
to the stellar rotation period at the midpoint of this interval.
The obliquity can be aligned before rapid orbital decay when $Q'_{10, 10days} \lesssim 10^{-2} Q'_e$. The stellar rotational period increases significantly during this phase.}
\label{f:tidepar}
\end{figure}

The results are shown in figure \ref{f:tidepar}. The color of each circle encodes the ratio of the timescales. The size of each circle encodes the stellar spin period during the interval between obliquity damping and orbital decay. When $Q'_{10}$ is large, the tidal component does not dominate and the obliquity is damped on a similar timescale as orbital decay. The spin period in this case is not increased significantly. When $Q'_{10}$ is relatively small, the star tends to be driven to $180^\circ$ obliquity and slowed down due to the equilibrium tide, before switching to the prograde state (similar to the case of the dotted purple line in Figure \ref{f:tide}). In this case, the alignment timescale can reach $\approx$85\% that of the orbital decay timescale. When $Q'_{10}$ is small and the initial stellar rotation is fast, the obliquity stalls at $90^\circ$ before spin-orbit alignment, and the alignment timescale can reach 10-30\% that of the orbital decay timescale. The stellar spin periods are increased significantly in the last two cases.

In summary, we have shown that the combination of the $(1,0)$ tidal component, the equilibrium tide, and magnetic braking can align the stellar spin before orbital decay, even from an initially retrograde configuration.  The stellar obliquity can be aligned long before orbital decay ($t_{align}/t_{a} \lesssim 0.1-0.3$) when $Q'_{10, 10~{\rm days}}/Q'_e \lesssim 10^{-3}$ and $\Omega_s \gtrsim 0.3 \Omega$.  The stellar obliquity can also be aligned prior to orbital decay when $Q'_{10, 10~{\rm days}} / Q'_e$ is as large as $10^{-2}$, but only for a relatively shorter interval ($t_{align}/t_{a} \lesssim 0.9$).  Systems that have been aligned in this manner would tend to have longer-than-usual stellar rotation periods.

In this scenario, the stellar obliquity spends more time near $90^\circ$ and $180^\circ$ compared with intermediate values of the obliquity. Thus, if this scenario is correct for explaining the spin-orbit alignment of the closest-in giant planets, then for systems with somewhat larger star-planet separations (where the tidal timescales are longer), there should be a preference for obliquities near $90^\circ$ or $180^\circ$. In this sense we agree with the conclusions of \citet{Rogers13b} and \citet{Lai12} that the dominance of the $(1,0)$ tidal component would produce substantial fractions of retrograde and perpendicular systems, but perhaps only for systems with relatively distant orbits.

\section{Conclusions and Discussions}
\label{s:diss}

\citet{Mazeh15} pointed out that for a population of host stars of transiting planets, the observed amplitude of photometric variability should be related to the degree of spin-orbit alignment.  They detected enhanced variability among the relatively cool host stars of {\it Kepler} transiting planets, indicating spin-orbit alignment. We have extended this study with statistical tests suggesting that this tendency weakens as the planet's orbital period grows longer.  Our statistical tests also indicate that the relatively low photometric variability of stars with long-period planets is not purely due to selection effects. In particular, using the geometric interpretation, we find that the evidence for alignment becomes weaker for systems with an innermost planet period $\gtrsim$10~days, and is consistent with nearly random alignment for longer orbital periods ($\gtrsim$30~days).

In addition, we found no evidence for any differences in the photometric variability of the multi-transiting systems and the short-period single-transiting systems. In contrast, the long-period single-transiting systems show reduced variability characteristic of a broad obliquity distribution. The dependence of the star's obliquity upon the current orbital configuration of the planetary system is seemingly at odds with theories in which obliquities are purely the outcome of star-disk processes in the ancient past \citep{Batygin12, Bate10, Thies11, Fielding15}, or the tidal effect of a planet that no longer exists \citep{Matsakos15}.

The overall dependence of the spin-orbit misalignment on the orbital period is at least qualitatively consistent with the scenario proposed by \citet{Winn10}, where the stellar obliquity is initially random, and is re-aligned due to tidal effects. We have also confirmed that with the tidal component identified by \citet{Lai12}, it is possible for a close-in giant planet to re-align the stellar obliquity long before rapid orbital decay from a retrograde state without assuming that only the outer layer of the star is re-aligned. Such alignment process also leads to a significant increase in the stellar rotational period, which can help identify the systems aligned due to the tidal component. If this scenario is widely applicable then stars with relatively wide-orbiting planets (and long tidal timescales) should often be found with obliquities of $90^\circ$ and $180^\circ$. 

However, a major quantitative problem with tidal theories is that the correlation between photometric variability and orbital period seems to extend to longer orbital periods ($\approx$30~days) than can easily be accounted for by tidal effects. Based on current estimates of the relevant tidal timescales, it seems difficult to re-align the star and planet when the innermost planet has a period longer than about 5 days. Other mechanisms may be required to align stars with periods $5-30$ days, or to preferentially produce larger misalignments for wider-orbiting planets. We briefly review some of the proposed mechanisms. Scattering by neighboring stars does preferentially produce larger misalignment at larger planet-star distances \citep[e.g., ][]{Li15}, however, the scattering effect is weak for the inner planets within $\sim$5~AU. Warping of the disk at close distance can be difficult, and the nodal precession of the disk with large warps may destroy the disk.  Another possibility is scattering within a system of close-in planets; however, it has proven difficult to excite the inclination to large enough values ($\gtrsim$5$^\circ$) \citep{Petrovich14}. The magnetic realignment mechanism \citep{Spalding15} depends on the inner truncation radius of the protoplanetary disk, which may be related to the orbital distance of the innermost planet; this would lead in principle to a dependence of spin-orbit misalignment on orbital distance.  In the ``early stellar ingestion'' theory of \citep{Matsakos15}, the agent of alignment is a giant planet that no longer exists, and it is difficult to know whether this scenario would lead to any relationship between spin-orbit misalignment and the orbital distances of surviving planets. Undoubtedly, the story of stellar obliquities is still missing some important chapters.

\acknowledgements The authors thank the anonymous referee for a constructive critique of the manuscript; Hyung Suk Tak, Kevin Schlaufman and Molei Tao for advice on the statistical methods; Dong Lai for helpful discussions on the tidal models; and Smadar Naoz, Konstantin Batygin and Bekki Dawson for helpful discussions on misalignment mechanisms. GL is grateful for the helpful discussions and lectures at the Summer School in Statistics for Astronomers XI at the Penn State University. JNW is grateful to the NASA Origins program for financial support (grant NNX11AG85G).

\bibliographystyle{hapj}
\bibliography{msref.bib}

\begin{thebibliography}{55}
\expandafter\ifx\csname natexlab\endcsname\relax\def\natexlab#1{#1}\fi

\bibitem[{{Akeson} {et~al.}(2013){Akeson}, {Chen}, {Ciardi}, {Crane}, {Good},
  {Harbut}, {Jackson}, {Kane}, {Laity}, {Leifer}, {Lynn}, {McElroy}, {Papin},
  {Plavchan}, {Ram{\'{\i}}rez}, {Rey}, {von Braun}, {Wittman}, {Abajian},
  {Ali}, {Beichman}, {Beekley}, {Berriman}, {Berukoff}, {Bryden}, {Chan},
  {Groom}, {Lau}, {Payne}, {Regelson}, {Saucedo}, {Schmitz}, {Stauffer},
  {Wyatt}, \& {Zhang}}]{Akeson13}
{Akeson}, R.~L. {et~al.} 2013, \pasp, 125, 989, 1307.2944

\bibitem[{{Albrecht} {et~al.}(2012){Albrecht}, {Winn}, {Johnson}, {Howard},
  {Marcy}, {Butler}, {Arriagada}, {Crane}, {Shectman}, {Thompson}, {Hirano},
  {Bakos}, \& {Hartman}}]{Albrecht12}
{Albrecht}, S. {et~al.} 2012, \apj, 757, 18, 1206.6105

\bibitem[{{Angus} {et~al.}(2015){Angus}, {Aigrain}, {Foreman-Mackey}, \&
  {McQuillan}}]{Angus15}
{Angus}, R., {Aigrain}, S., {Foreman-Mackey}, D., \& {McQuillan}, A. 2015,
  \mnras, 450, 1787, 1502.06965

\bibitem[{{Barker} \& {Ogilvie}(2009)}]{Barker09}
{Barker}, A.~J., \& {Ogilvie}, G.~I. 2009, in IAU Symposium, Vol. 259, IAU
  Symposium, ed. K.~G. {Strassmeier}, A.~G. {Kosovichev}, \& J.~E. {Beckman},
  295--302, 0902.4554

\bibitem[{{Bate} {et~al.}(2010){Bate}, {Lodato}, \& {Pringle}}]{Bate10}
{Bate}, M.~R., {Lodato}, G., \& {Pringle}, J.~E. 2010, \mnras, 401, 1505,
  0909.4255

\bibitem[{{Batygin}(2012)}]{Batygin12}
{Batygin}, K. 2012, \nat, 491, 418

\bibitem[{{Chatterjee} {et~al.}(2008){Chatterjee}, {Ford}, {Matsumura}, \&
  {Rasio}}]{Chatterjee08}
{Chatterjee}, S., {Ford}, E.~B., {Matsumura}, S., \& {Rasio}, F.~A. 2008, \apj,
  686, 580, arXiv:astro-ph/0703166

\bibitem[{{Dawson}(2014)}]{Dawson14}
{Dawson}, R.~I. 2014, \apjl, 790, L31, 1405.1735

\bibitem[{{Fabrycky} \& {Tremaine}(2007)}]{Fabrycky07}
{Fabrycky}, D., \& {Tremaine}, S. 2007, \apj, 669, 1298, 0705.4285

\bibitem[{{Fabrycky} \& {Winn}(2009)}]{FabryckyWinn09}
{Fabrycky}, D.~C., \& {Winn}, J.~N. 2009, \apj, 696, 1230, 0902.0737

\bibitem[{{Favier} {et~al.}(2014){Favier}, {Barker}, {Baruteau}, \&
  {Ogilvie}}]{Favier14}
{Favier}, B., {Barker}, A.~J., {Baruteau}, C., \& {Ogilvie}, G.~I. 2014,
  \mnras, 439, 845, 1401.0643

\bibitem[{{Fielding} {et~al.}(2015){Fielding}, {McKee}, {Socrates},
  {Cunningham}, \& {Klein}}]{Fielding15}
{Fielding}, D.~B., {McKee}, C.~F., {Socrates}, A., {Cunningham}, A.~J., \&
  {Klein}, R.~I. 2015, \mnras, 450, 3306, 1409.5148

\bibitem[{{Foucart} \& {Lai}(2011)}]{Foucart11}
{Foucart}, F., \& {Lai}, D. 2011, \mnras, 412, 2799, 1009.3233

\bibitem[{{H{\'e}brard} {et~al.}(2008){H{\'e}brard}, {Bouchy}, {Pont},
  {Loeillet}, {Rabus}, {Bonfils}, {Moutou}, {Boisse}, {Delfosse}, {Desort},
  {Eggenberger}, {Ehrenreich}, {Forveille}, {Lagrange}, {Lovis}, {Mayor},
  {Pepe}, {Perrier}, {Queloz}, {Santos}, {S{\'e}gransan}, {Udry}, \&
  {Vidal-Madjar}}]{Hebrard08}
{H{\'e}brard}, G. {et~al.} 2008, \aap, 488, 763, 0806.0719

\bibitem[{{H{\'e}brard} {et~al.}(2011){H{\'e}brard}, {Ehrenreich}, {Bouchy},
  {Delfosse}, {Moutou}, {Arnold}, {Boisse}, {Bonfils}, {D{\'{\i}}az},
  {Eggenberger}, {Forveille}, {Lagrange}, {Lovis}, {Pepe}, {Perrier}, {Queloz},
  {Santerne}, {Santos}, {S{\'e}gransan}, {Udry}, \& {Vidal-Madjar}}]{Hebrard11}
------. 2011, \aap, 527, L11, 1101.5009

\bibitem[{{Hirano} {et~al.}(2011){Hirano}, {Narita}, {Sato}, {Winn}, {Aoki},
  {Tamura}, {Taruya}, \& {Suto}}]{Hirano11}
{Hirano}, T., {Narita}, N., {Sato}, B., {Winn}, J.~N., {Aoki}, W., {Tamura},
  M., {Taruya}, A., \& {Suto}, Y. 2011, \pasj, 63, L57, 1108.4493

\bibitem[{{Huber} {et~al.}(2013){Huber}, {Carter}, {Barbieri}, {Miglio},
  {Deck}, {Fabrycky}, {Montet}, {Buchhave}, {Chaplin}, {Hekker},
  {Montalb{\'a}n}, {Sanchis-Ojeda}, {Basu}, {Bedding}, {Campante},
  {Christensen-Dalsgaard}, {Elsworth}, {Stello}, {Arentoft}, {Ford},
  {Gilliland}, {Handberg}, {Howard}, {Isaacson}, {Johnson}, {Karoff},
  {Kawaler}, {Kjeldsen}, {Latham}, {Lund}, {Lundkvist}, {Marcy}, {Metcalfe},
  {Silva Aguirre}, \& {Winn}}]{Huber13}
{Huber}, D. {et~al.} 2013, Science, 342, 331, 1310.4503

\bibitem[{{Huber} {et~al.}(2014){Huber}, {Silva Aguirre}, {Matthews},
  {Pinsonneault}, {Gaidos}, {Garc{\'{\i}}a}, {Hekker}, {Mathur}, {Mosser},
  {Torres}, {Bastien}, {Basu}, {Bedding}, {Chaplin}, {Demory}, {Fleming},
  {Guo}, {Mann}, {Rowe}, {Serenelli}, {Smith}, \& {Stello}}]{Huber14}
------. 2014, \apjs, 211, 2, 1312.0662

\bibitem[{{Jackson} \& {Jeffries}(2013)}]{JacksonJeffries13}
{Jackson}, R.~J., \& {Jeffries}, R.~D. 2013, \mnras, 431, 1883, 1302.4202

\bibitem[{{Lai}(2012)}]{Lai12}
{Lai}, D. 2012, \mnras, 423, 486, 1109.4703

\bibitem[{{Lai} {et~al.}(2011){Lai}, {Foucart}, \& {Lin}}]{Lai11}
{Lai}, D., {Foucart}, F., \& {Lin}, D.~N.~C. 2011, \mnras, 412, 2790, 1008.3148

\bibitem[{{Li} \& {Adams}(2015)}]{Li15}
{Li}, G., \& {Adams}, F.~C. 2015, \mnras, 448, 344, 1501.00911

\bibitem[{{Li} {et~al.}(2014{\natexlab{a}}){Li}, {Naoz}, {Kocsis}, \&
  {Loeb}}]{Li14}
{Li}, G., {Naoz}, S., {Kocsis}, B., \& {Loeb}, A. 2014{\natexlab{a}}, \apj,
  785, 116, 1310.6044

\bibitem[{{Li} {et~al.}(2014{\natexlab{b}}){Li}, {Naoz}, {Valsecchi},
  {Johnson}, \& {Rasio}}]{Li14_56}
{Li}, G., {Naoz}, S., {Valsecchi}, F., {Johnson}, J.~A., \& {Rasio}, F.~A.
  2014{\natexlab{b}}, \apj, 794, 131, 1407.2249

\bibitem[{{Matsakos} \& {K{\"o}nigl}(2015)}]{Matsakos15}
{Matsakos}, T., \& {K{\"o}nigl}, A. 2015, \apjl, 809, L20, 1507.07967

\bibitem[{{Mazeh} {et~al.}(2015){Mazeh}, {Perets}, {McQuillan}, \&
  {Goldstein}}]{Mazeh15}
{Mazeh}, T., {Perets}, H.~B., {McQuillan}, A., \& {Goldstein}, E.~S. 2015,
  \apj, 801, 3, 1501.01288

\bibitem[{{McQuillan} {et~al.}(2013){McQuillan}, {Mazeh}, \&
  {Aigrain}}]{McQuillan13}
{McQuillan}, A., {Mazeh}, T., \& {Aigrain}, S. 2013, \apjl, 775, L11, 1308.1845

\bibitem[{{McQuillan} {et~al.}(2014){McQuillan}, {Mazeh}, \&
  {Aigrain}}]{McQuillan14}
------. 2014, \apjs, 211, 24, 1402.5694

\bibitem[{{Morton} \& {Winn}(2014)}]{Morton14}
{Morton}, T.~D., \& {Winn}, J.~N. 2014, \apj, 796, 47, 1408.6606

\bibitem[{{Nagasawa} {et~al.}(2008){Nagasawa}, {Ida}, \& {Bessho}}]{Nagasawa08}
{Nagasawa}, M., {Ida}, S., \& {Bessho}, T. 2008, \apj, 678, 498, 0801.1368

\bibitem[{{Naoz} {et~al.}(2011){Naoz}, {Farr}, {Lithwick}, {Rasio}, \&
  {Teyssandier}}]{Naoz11}
{Naoz}, S., {Farr}, W.~M., {Lithwick}, Y., {Rasio}, F.~A., \& {Teyssandier}, J.
  2011, \nat, 473, 187, 1011.2501

\bibitem[{{Naoz} {et~al.}(2012){Naoz}, {Farr}, \& {Rasio}}]{Naoz12}
{Naoz}, S., {Farr}, W.~M., \& {Rasio}, F.~A. 2012, \apjl, 754, L36, 1206.3529

\bibitem[{{Narita} {et~al.}(2009){Narita}, {Sato}, {Hirano}, \&
  {Tamura}}]{Narita09}
{Narita}, N., {Sato}, B., {Hirano}, T., \& {Tamura}, M. 2009, \pasj, 61, L35,
  0908.1673

\bibitem[{{Ogilvie}(2005)}]{Ogilvie05}
{Ogilvie}, G.~I. 2005, Journal of Fluid Mechanics, 543, 19, astro-ph/0506450

\bibitem[{{Ogilvie}(2013)}]{Ogilvie13}
------. 2013, \mnras, 429, 613, 1211.0837

\bibitem[{{Ogilvie}(2014)}]{Ogilvie14}
------. 2014, \araa, 52, 171, 1406.2207

\bibitem[{{Ogilvie} \& {Lin}(2007)}]{Ogilvie07}
{Ogilvie}, G.~I., \& {Lin}, D.~N.~C. 2007, \apj, 661, 1180, astro-ph/0702492

\bibitem[{{Petrovich} {et~al.}(2014){Petrovich}, {Tremaine}, \&
  {Rafikov}}]{Petrovich14}
{Petrovich}, C., {Tremaine}, S., \& {Rafikov}, R. 2014, \apj, 786, 101,
  1401.4457

\bibitem[{{Pont}(2009)}]{Pont09}
{Pont}, F. 2009, \mnras, 396, 1789, 0812.1463

\bibitem[{{Rogers} \& {Lin}(2013)}]{Rogers13b}
{Rogers}, T.~M., \& {Lin}, D.~N.~C. 2013, \apjl, 769, L10, 1304.4148

\bibitem[{{Rogers} {et~al.}(2012){Rogers}, {Lin}, \& {Lau}}]{Rogers12}
{Rogers}, T.~M., {Lin}, D.~N.~C., \& {Lau}, H.~H.~B. 2012, \apjl, 758, L6,
  1209.2435

\bibitem[{{Rogers} {et~al.}(2013){Rogers}, {Lin}, {McElwaine}, \&
  {Lau}}]{Rogers13}
{Rogers}, T.~M., {Lin}, D.~N.~C., {McElwaine}, J.~N., \& {Lau}, H.~H.~B. 2013,
  \apj, 772, 21, 1306.3262

\bibitem[{{Schlaufman}(2010)}]{Schlaufman10}
{Schlaufman}, K.~C. 2010, \apj, 719, 602, 1006.2851

\bibitem[{{Spalding} \& {Batygin}(2014)}]{Spalding14}
{Spalding}, C., \& {Batygin}, K. 2014, \apj, 790, 42, 1406.4183

\bibitem[{{Spalding} \& {Batygin}(2015)}]{Spalding15}
------. 2015, \apj, 811, 82, 1508.02365

\bibitem[{{Teitler} \& {K{\"o}nigl}(2014)}]{Teitler14}
{Teitler}, S., \& {K{\"o}nigl}, A. 2014, \apj, 786, 139, 1403.5860

\bibitem[{{Thies} {et~al.}(2011){Thies}, {Kroupa}, {Goodwin}, {Stamatellos}, \&
  {Whitworth}}]{Thies11}
{Thies}, I., {Kroupa}, P., {Goodwin}, S.~P., {Stamatellos}, D., \& {Whitworth},
  A.~P. 2011, \mnras, 417, 1817, 1107.2113

\bibitem[{{Tremaine}(1989)}]{Tremaine89}
{Tremaine}, S. 1989, in Dynamics of Astrophysical Discs, ed. J.~A. {Sellwood},
  231--238

\bibitem[{{Triaud} {et~al.}(2010){Triaud}, {Collier Cameron}, {Queloz},
  {Anderson}, {Gillon}, {Hebb}, {Hellier}, {Loeillet}, {Maxted}, {Mayor},
  {Pepe}, {Pollacco}, {S{\'e}gransan}, {Smalley}, {Udry}, {West}, \&
  {Wheatley}}]{Triaud10}
{Triaud}, A.~H.~M.~J. {et~al.} 2010, \aap, 524, A25, 1008.2353

\bibitem[{{Valsecchi} \& {Rasio}(2014)}]{Valsecchi14}
{Valsecchi}, F., \& {Rasio}, F.~A. 2014, \apj, 786, 102, 1402.3857

\bibitem[{{Winn} {et~al.}(2010){Winn}, {Fabrycky}, {Albrecht}, \&
  {Johnson}}]{Winn10}
{Winn}, J.~N., {Fabrycky}, D., {Albrecht}, S., \& {Johnson}, J.~A. 2010, \apjl,
  718, L145, 1006.4161

\bibitem[{{Winn} \& {Fabrycky}(2015)}]{Winn14}
{Winn}, J.~N., \& {Fabrycky}, D.~C. 2015, \araa, 53, 409, 1410.4199

\bibitem[{{Winn} {et~al.}(2009){Winn}, {Johnson}, {Fabrycky}, {Howard},
  {Marcy}, {Narita}, {Crossfield}, {Suto}, {Turner}, {Esquerdo}, \&
  {Holman}}]{Winn09}
{Winn}, J.~N. {et~al.} 2009, \apj, 700, 302, 0902.3461

\bibitem[{{Wu} \& {Lithwick}(2011)}]{Wu11}
{Wu}, Y., \& {Lithwick}, Y. 2011, \apj, 735, 109, 1012.3475

\bibitem[{{Xue} {et~al.}(2014){Xue}, {Suto}, {Taruya}, {Hirano}, {Fujii}, \&
  {Masuda}}]{Xue14}
{Xue}, Y., {Suto}, Y., {Taruya}, A., {Hirano}, T., {Fujii}, Y., \& {Masuda}, K.
  2014, \apj, 784, 66, 1401.5876

\end{thebibliography}

\end{document}